\documentclass{article}
\usepackage{romp}
\usepackage{amsmath,amsthm,amssymb,amsfonts}
\usepackage{graphicx}
\usepackage{url}
\usepackage{cancel}
\usepackage[english]{babel}

\newcommand{\moosewasy}{\fontencoding{U}\fontfamily{wasy}\selectfont}
\newcommand{\hexagon}{\mbox{\moosewasy\symbol{55}}}

\title{Bell polynomials in the series expansions of~the~Ising~model}
\author{  Grzegorz Siudem\thanks{The work has been supported by the National Science Centre of Poland (Narodowe Centrum Nauki, NCN) under grant no.  2012/05/E/ST2/02300.} and Agata Fronczak\thanks{The work has been supported by the National Science Centre of Poland (Narodowe Centrum Nauki, NCN) under grant no.  2015/18/E/ST2/00560.}\\ Faculty of Physics, Warsaw University of Technology,\\ ul.~Koszykowa 75, PL-00-662, Warsaw, Poland  }

\begin{document}

\maketitle
\begin{abstract}
Through applying Bell polynomials to the integral representation of the free energy of the Ising model for the triangular and  hexagonal lattices we obtain the exact combinatorial formulas for the number of spin configurations  at a given energy (i.e. low-temperature series expansion of the partition function or, alternatively, the number of states). We also generalize this approach to the wider class of the (chequered) Utiyama graphs. Apart from the presented exact formulas, our technique allows one to establish the correspondence between the perfect gas of clusters  and the Ising model on the lattices which have positive coefficients in the low-temperature expansion (e.g. square lattice, hexagonal lattice). However it is not always the case -- we present that for the triangular lattice the coefficients could be negative and the perfect gas of clusters  interpretation is problematic.
\end{abstract}


\section{Introduction}


In statistical physics, the research in the field of lattice models (i.e. simplified graph-based models of the real magnetic solids, which allows to better understand magnetic phase transitions), has almost $100$ years history \cite{ Brush1967,Niss2005}. It started with the works of Lenz \cite{Lenz1920} and his student Ising \cite{Ising1925}, . Therefore the name of the most known lattice model -- the Ising model could  seem unfair\footnote{Brush \cite{Brush1967} suggested to use name Ising-Lenz, but it seems not to be the mainstream nomenclature and most authors use simply Ising model. McCoy and Wu argued in their book \cite{ McCoy2010} (page 2. footnote 1), that however Lenz first wrote about the model in 1920 \cite{Lenz1920}, but he did not obtain any of its properties. However, Lenz encouraged Ising for his research (see quotes  collected by Brush \cite{Brush1967}).} to Lenz, who was the originator of the idea in 1920. In 1925 Ising  solved the one-dimensional case of the model and his result was not very promising \cite{Ising1925} -- he found that in the case of one-dimensional chain of spins there is no phase transition at all.  However, later efforts of mathematicians and physicists revealed the beauty and richness of the lattice models far greater than the very first one-dimensional Ising approach. An extensive historical discussion of the topic can be found in \cite{Brush1967,Niss2005}.

Usually by lattice models one means the idealization of the real magnetic crystal as a set of spins (in the simplified case as  variables $s_i=\pm 1$) put into vertices $i\in V_\mathcal{G}$ of the graph $\mathcal{G}=(V_\mathcal{G},E_\mathcal{G})$, where $V_\mathcal{G}$ is a set of graph's vertices and $E_\mathcal{G}$ is the set of its edges. For the Ising case, the energy of the system is given as
 \begin{equation}
 \label{eq:hamiltonianIsinga}	H((s_i))=-J\sum_{\{ l,\,p\}\in E_\mathcal{G}}s_ls_p,
 \end{equation}
where $-J$ is the energy of a pair of parallel spins  and summation is taken over all edges in $\mathcal{G}$ i.e. over the set  $E_\mathcal{G}$  of unordered pairs of connected vertices ($\{ l,\,p\}=\{ p,\,l\}$ for every $\{ l,\,p\}\in E_\mathcal{G}$). Let us note that in the further discussion for brevity we write $E=|E_\mathcal{G}|$ and $V=|V_\mathcal{G}|$ for the numbers of graph's edges and vertices respectively.
 
 The Ising model is definitely  one of the most impressive examples of a problem which is very easy to define and very hard to solve. Despite the years of efforts of physicists and mathematicians (however the problem arises from statstical physics it has also very significant immpact of the combinatorics, see \cite{Chelkak2017,Cipra1987,Loebl2010}), there are still fundamental and open questions about properties of the Ising model for the lattices more complex than a one-dimensional chain  (see chapter 8.1.4 in~\cite{Sethna2011} or chapters 10.1 and~10.4 in \cite{McCoy2010} for a wide discussion of such problems).  Despite those questions by {\it solving} the model one usually mean finding the {\it compact} formula for its partition function  $\mathfrak{Z}^{\mathcal{G}}_V$ (or free energy  $\mathfrak{F}^{\mathcal{G}}_V$)
  \begin{equation} \label{eq:partitionfunctionG}	
  \mathfrak{Z}^{\mathcal{G}}_V=\sum_{(s_i)\in\{-1,\,1\}^V}\exp\left[-\beta H((s_i))\right],\;\;\;\;\mathfrak{F}^{\mathcal{G}}_V=-\frac{1}{\beta}\ln \mathfrak{Z}^{\mathcal{G}}_V,
 \end{equation}
 where $\beta=1/(k_BT)$ is a standard thermal factor. As we mentioned above, the main goal of the investigation of the lattice models is to better understand the nature of the phase transitions, which  is a background of the importance of the partition function  $\mathfrak{Z}^{\mathcal{G}}_V$ (or, equivalently, free energy  $\mathfrak{F}^{\mathcal{G}}_V$). It follows in a straightforward manner from Lee-Yang theorem \cite{leeyang} that the phase transition in the model implies a non-analytical behavior of the partition function (free energy), thus there is no such transition for the finite graphs $\mathcal{G}$, where the function is just a polynomial. Therefore, it is natural to investigate infinite graphs, rather than their finite analogues. From the other hand, quantities given by Eq. (\ref{eq:partitionfunctionG}) diverge to infinity and this is the reason why one rather consider  free energy and partition function calculated per spin
   \begin{equation} \label{eq:partitionfunctionD}	
  Z^{\mathcal{G}}_V=\sqrt[V]{ \mathfrak{Z}^{\mathcal{G}}_V},\;\;\;\;F^{\mathcal{G}}_V=\frac{\mathfrak{F}^{\mathcal{G}}_V}{V},
 \end{equation}
which, in the limit of infinite lattice, gives so-called bulk case
    \begin{equation} \label{eq:partitionfunctionB}	
  \zeta_{\mathcal{G}}=\lim_{V\to \infty}Z^\mathcal{G}_V=\lim_{V\to \infty}\sqrt[V]{ \mathfrak{Z}^{\mathcal{G}}_V},\;\;\;\;\varphi_{\mathcal{G}}=\lim_{V\to\infty}F^\mathcal{G}_V=\lim_{V\to\infty}\frac{\mathfrak{F}^{\mathcal{G}}_V}{V}.
 \end{equation}
 \remark{Remark}{\label{rem:oznaczenia}
 As one see in Eqs. (\ref{eq:partitionfunctionG}, \ref{eq:partitionfunctionD}, \ref{eq:partitionfunctionB}),  there are three levels of description of the lattice model: considerations for finite graphs, which we denote with calligraphic fonts   ($\mathfrak{Z}_V^\mathcal{G},\,\mathfrak{F}_V^\mathcal{G},\,\mathfrak{g}_V^\mathcal{G},\,\mathfrak{a}_V^\mathcal{G}$), the density-like version, when one calculates free energy per spin, which is denoted with Greek letters and regular fonts with subscript $V$ i.e. size of the graph ($Z_V^\mathcal{G},\,F_V^\mathcal{G}\,g_V^\mathcal{G},\,a_V^\mathcal{G}$) and the (infinite) bulk case with the Greek letters and regular fonts with $\mathcal{G}$ in subscripts ($\zeta_\mathcal{G},\,\varphi_\mathcal{G},\,g_\mathcal{G},\,a_\mathcal{G}$). The meaning of coefficients $g$ and $a$ is presented in Eqs. (\ref{eq:serlev1}, \ref{eq:serlev2}, \ref{eq:serlev3}).}

 As already mentioned, the works on the Ising model started (by Ising himself \cite{Ising1925}) with the analysis in which graph $\mathcal{G}$ was a simple one-dimensional chain. The next step were the studies of two-dimensional planar graphs for which Kramers and Wannier \cite{kramers} introduced (known under their names) duality (of the different type of graphs), which allowed them to find the value of the critical temperature. Nevertheless, they did not prove that for such graphs the phase transition occurs, and it happened not earlier than in 1944 when Onsager solved \cite{Onsager1944} the  square lattice case. With his algebraic approach, he obtained the closed form for the bulk free energy in the following integral representation
 \begin{align}\label{eq:phiS}
-\beta\varphi_\square=&\ln 2 +\frac{1}{8\pi^2}\int_0^{2\pi}d\theta_1\int_0^{2\pi}d\theta_2 \ln\left[\cosh^2(2\beta J)-\sinh(2\beta J)(\cos\theta_1+\cos\theta_2)\right].
\end{align}
This result opened new directions (for more details see \cite{ Brush1967,Niss2005}) in the research of the lattice models and probably the best proof of its value is a quotation from the letter\footnote{The quotation comes from Onsager's obituary in {\it Physics Today} \cite{Montroll1977}.} from Pauli, in which he calms down Casimir concerns about being cut off for so long from the scientific results from the Allied countries. He wrote: {\it  Nothing much of interest has happened except for Onsager's exact solution of the Two-Dimensional Ising 	Model}.

\begin{figure}[ht]
\centering{\includegraphics[scale=0.28]{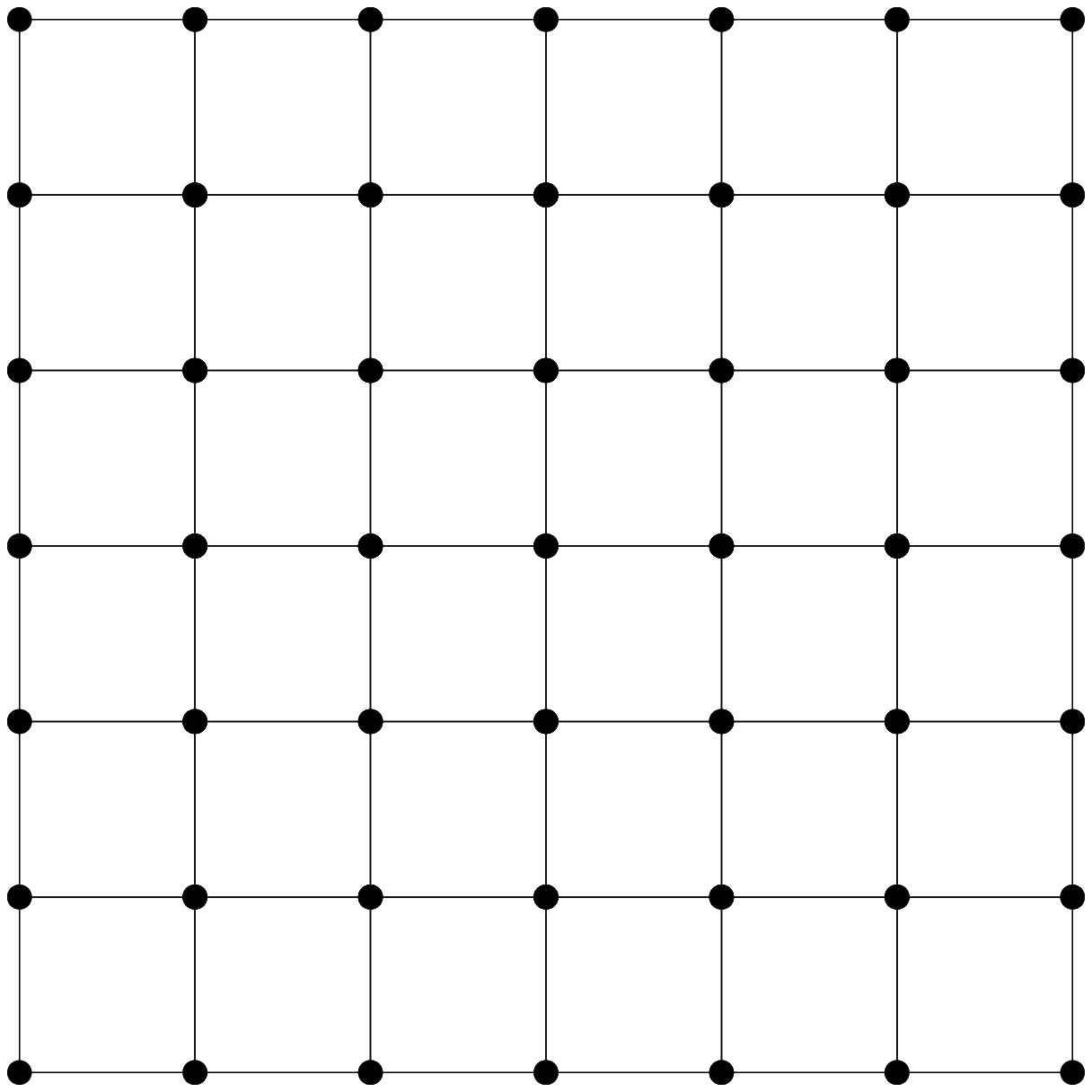}
\includegraphics[scale=0.33]{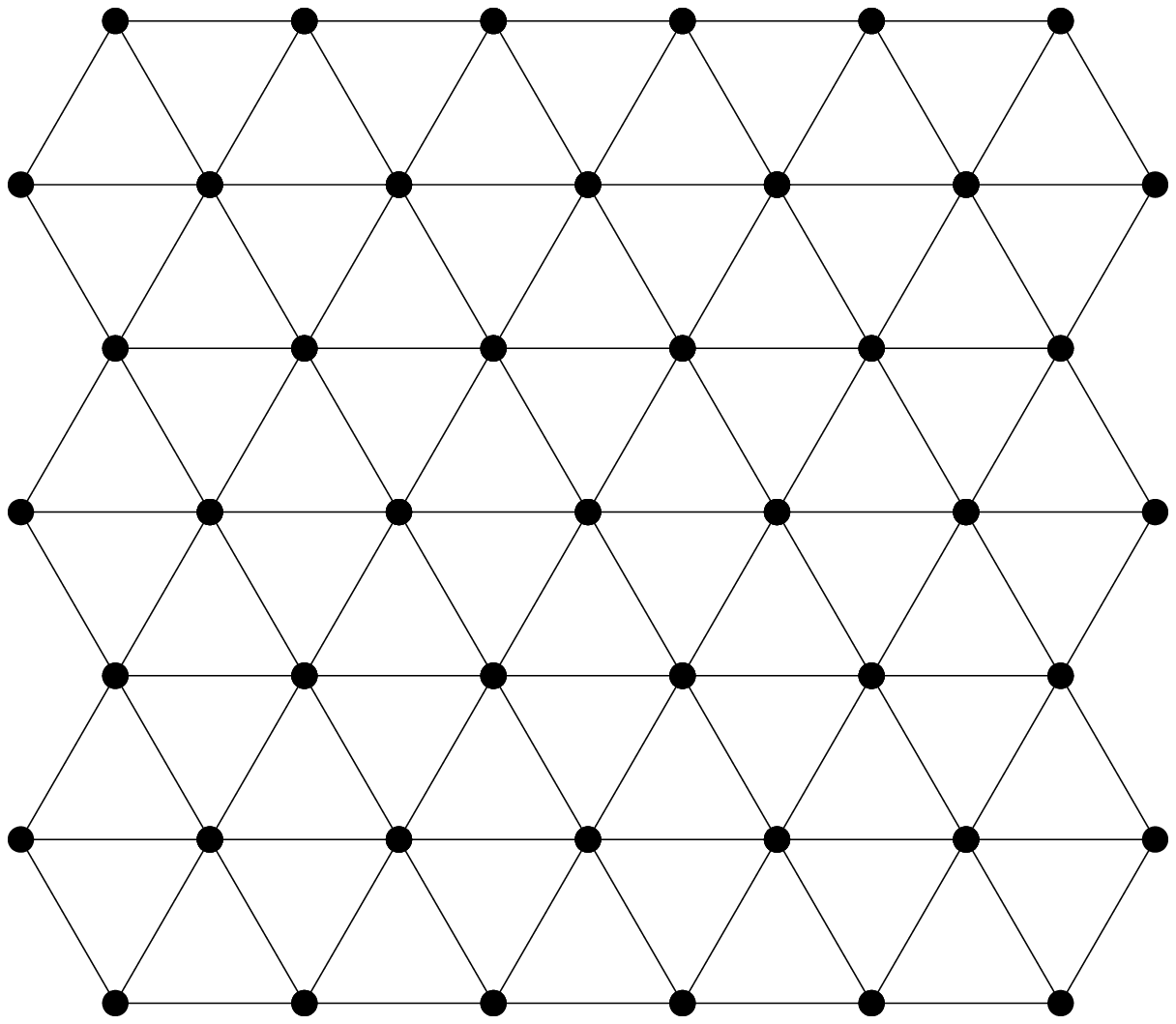} \includegraphics[scale=0.31]{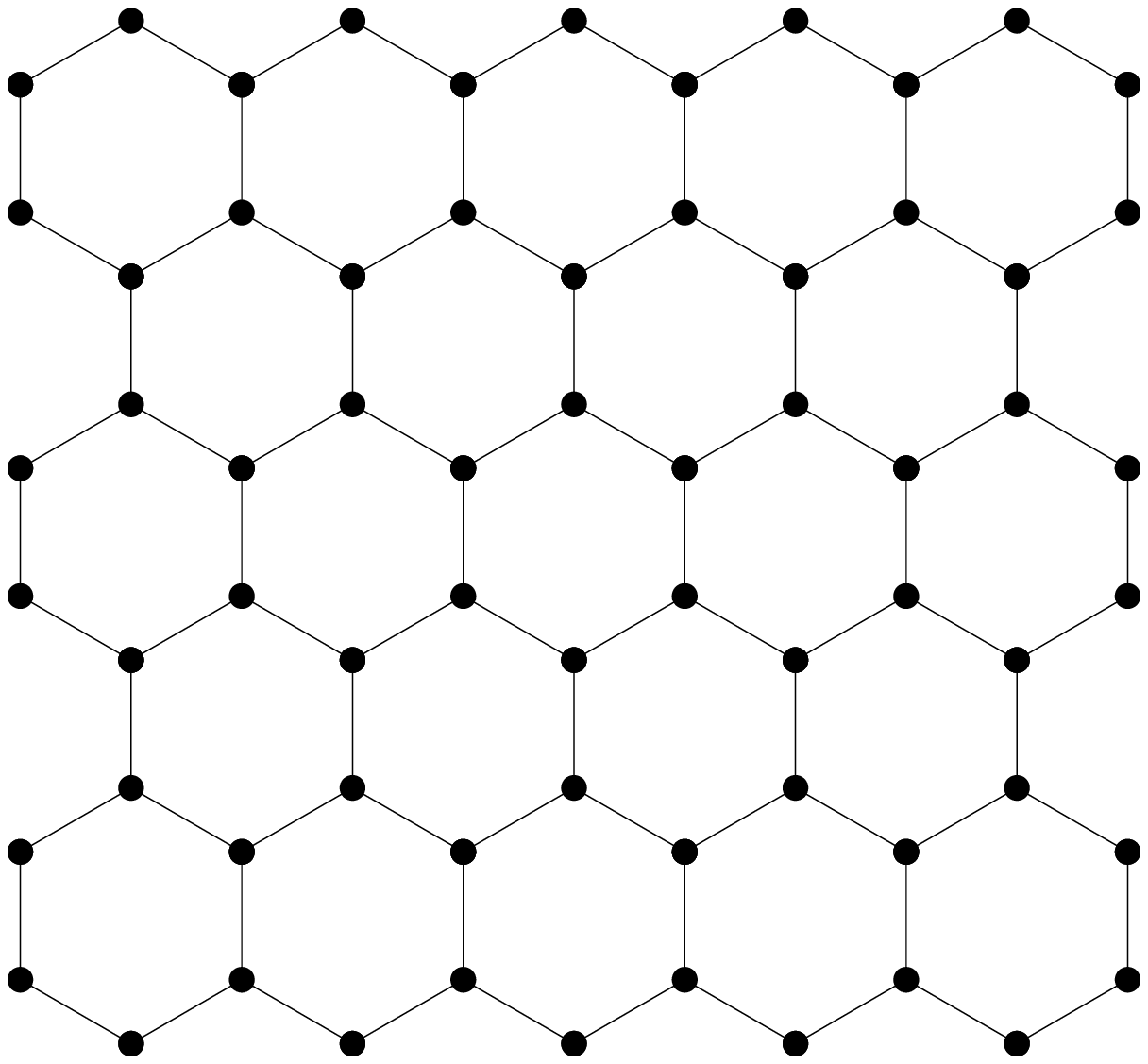}}
\caption{Considered lattices, from the left: square ($\mathcal{G}=\square$), triangular ($\mathcal{G}=\triangle$) and hexagonal ($\mathcal{G}={\hexagon}$).}
\label{fig:lattice}
\end{figure}

\begin{figure}[ht!]
\centering{\includegraphics[scale=0.35]{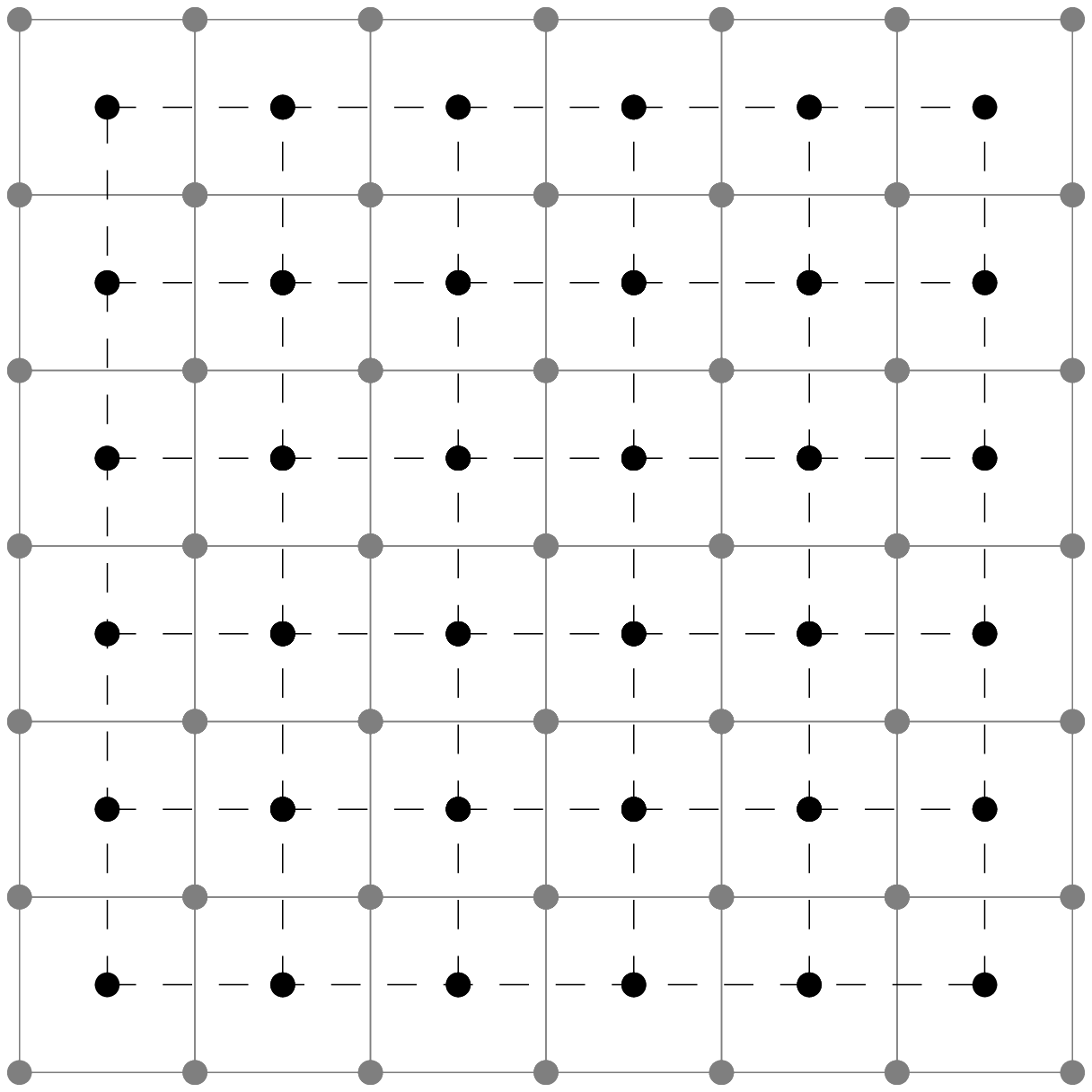}
\includegraphics[scale=0.38]{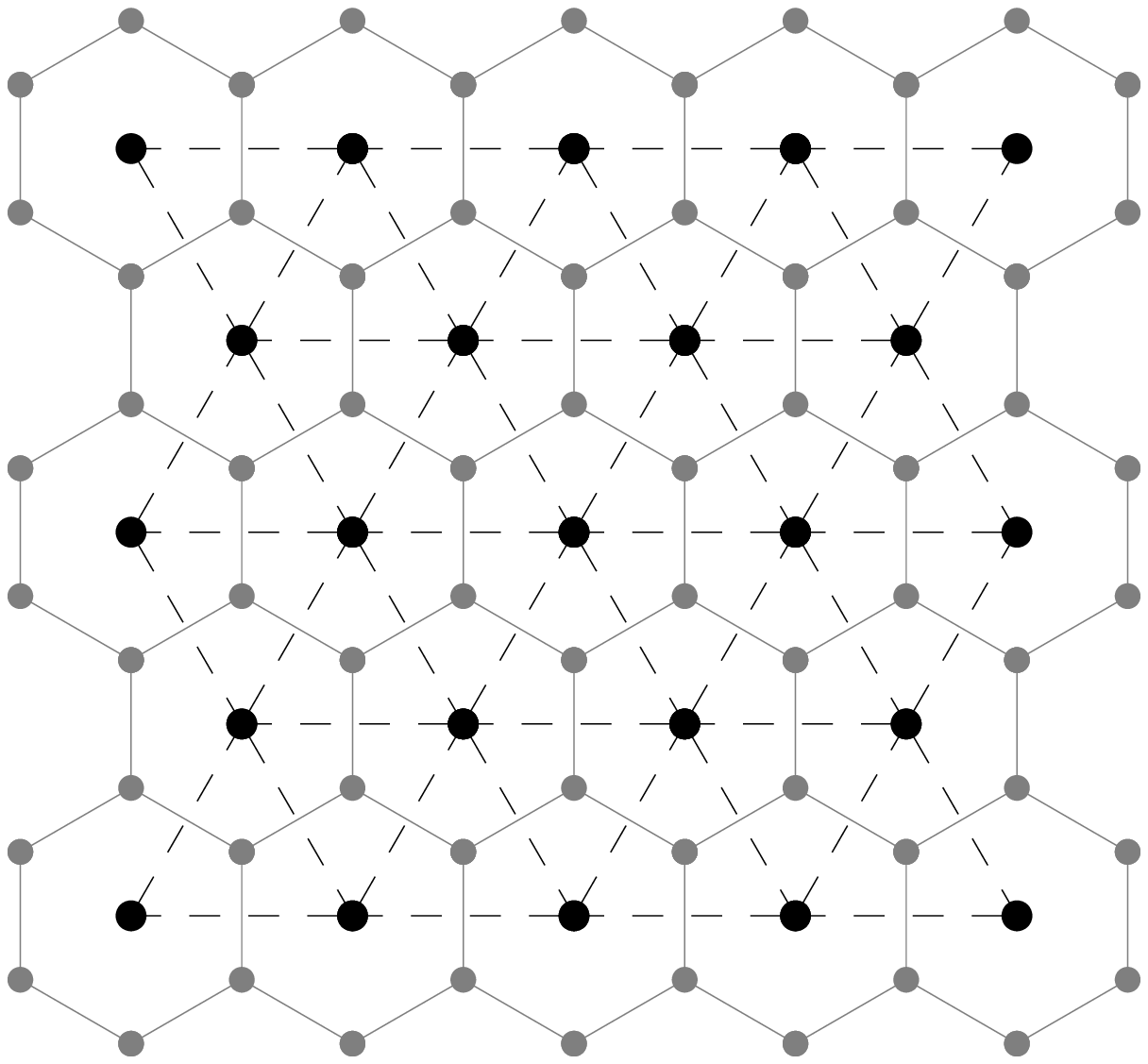}}

	\caption{\small Lattices from Fig. \ref{fig:lattice} with the Kramers-Wannier's dual-ones.}
	\label{fig:duality}
\end{figure}

Onsager's approach (more precisely and fairly Onsager-Kauffman\footnote{Let us pay reader's attention to, the interesting in that context, Onsager's letter to Kaufman \cite{Kaufman1995}, published for his 90th birthday \cite{Lebowitz1995}.  } \cite{Kaufman1949.2}, for the detailed discussion, see \cite{Niss2005}, p. 304) allowed Wannier \cite{wannier} to obtain an analogue of Eq. (\ref{eq:phiS}) for the triangular lattice. In this paper, we use an equivalent formula (for proof of the equivalence see appendix \ref{sec:formulae}) for the Ising model's bulk free energy for the triangular lattice in the form (see Eq. (127) in \cite{domb1})
\begin{align}
-\beta \varphi_\triangle=\ln 2+\frac{1}{8\pi^2}\int_0^{2\pi} d\theta_1& \int_0^{2\pi}
d\theta_2\;\ln\bigg\{\cosh^3(2\beta J)+\sinh^3(2\beta J)+\label{eq:phiT}\\&+\sinh(2\beta J)\bigg[\cos\theta_1+\cos\theta_2- \cos(\theta_1+\theta_2)\bigg]\bigg\}\nonumber.
\end{align}
Because of the fact, that triangular lattice is dual in the Kramers-Wannier sense (for the detailed discussion see chapter 13.1 in \cite{McCoy2010}) to the hexagonal one (see Fig. \ref{fig:duality}),  one can transform Eq. (\ref{eq:phiT}) for the corresponding formula for the bulk free energy of the Ising model on the hexagonal lattice\footnote{It seems \cite{Baxter399,Lu2001}, that this formula firstly appears in \cite{Houtappel1950} as Eq. (96). Let us note that analogical formula, (131) in Domb's monograph \cite{domb1}, from which we adopt the notion, has a few misprints.}
\begin{align}
-\beta \varphi_{\hexagon}&=\frac{3}{4}\ln 2+\frac{1}{16\pi^2}\int_0^{2\pi} d\theta_1 \int_0^{2\pi}
d\theta_2\times\label{eq:phiH}\\
\times &\ln\bigg\{\cosh^3(2\beta J)+1-\sinh^2(2\beta J)\bigg[\cos\theta_1+\cos\theta_2+ \cos(\theta_1+\theta_2)\bigg]\bigg\}\nonumber.
\end{align}
The integral representation of the free energies $\varphi_\mathcal{G}$ in the form of Eqs. (\ref{eq:phiS}, \ref{eq:phiT}, \ref{eq:phiH}) will be the starting point in our series expansions of the partition functions $\zeta_\mathcal{G},\,Z_V^\mathcal{G},\, \mathfrak{Z}_V^\mathcal{G}$  and other levels of free energies $F_V^\mathcal{G},\, \mathfrak{F}_V^\mathcal{G}$. Furthermore, 
the similarity of Eqs. (\ref{eq:phiS}, \ref{eq:phiT}, \ref{eq:phiH})  suggests that they are special cases of a one  general rule. In fact, all of them are examples of chequered lattices (see \cite{Utiyama1951} or  sec. 3.5.4 {\it (iv)} in  \cite{ domb1}) which we introduce in sec. \ref{sec:utiyama} and analyse in sec. \ref{sec:utiyamaapplication}. Bell polynomials play the crucial role in our considerations  (for details see sec. \ref{sec:toolsandmethods}), thus, in the following paper, we refer to the discussed  technique as {\it Bell polynomials' approach}.

The series expansions are typical technique used in statistical physics. They were widely applied, in particular, in the case of the lattice models (see sec. II in \cite{McCoy2010} and \cite{Beale1996,domb2,AF1,AF2,AFPF,Hen2018,GS,GSAFPF,Viswanathan2015,Zhou2018}). In this work we concentrate mainly on the low-temperature expansion and its dual (in sense of Kramers-Wannier) counterpart i.e. the high-temperature expansion (for details see  sec \ref{sec:variables} and mentioned above \cite{domb1,domb2,McCoy2010}. Let us introduce the  low-temperature  variable $x=e^{-2\beta J}$, which allows us to consider the series expansions of the introduced free energies and partition functions on three levels of description (see remark \ref{rem:oznaczenia} on page \pageref{rem:oznaczenia})  
\begin{align}
&\mathfrak{Z}_V^{\mathcal{G}}(x)=2x^{-\frac{E}{2}}\sum_{r=0}^{E}\mathfrak{g}_V^{\mathcal{G}}(r) x^r,\;&-\beta \mathfrak{F}_V^\mathcal{G}(x)=-\frac{E}{2}\ln x+\sum_{n=1}^\infty\frac{ \mathfrak{a}_V^\mathcal{G}(n)}{n!}x^n,\label{eq:serlev1}\\
&Z_V^\mathcal{G}(x)=2x^{-\frac{E}{2V}} \sum_{r=0}^{\infty}g_V^\mathcal{G}(r)x^r,\;	&-\beta F_V^\mathcal{G}(x)= -\frac{E}{2V}\ln x + \sum_{n=1}^\infty \frac{a_V^\mathcal{G}(n)}{n!}x^n,\label{eq:serlev2}\\
&\zeta_\mathcal{G}(x)=x^{-\frac{E}{2V}} \sum_{r=0}^{\infty}g_\mathcal{G}(r)x^r,\;&-\beta \varphi_\mathcal{G}(x)= -\frac{E}{2V}\ln x + \sum_{n=1}^\infty \frac{a_\mathcal{G}(n)}{n!}x^n.\label{eq:serlev3}
\end{align}

\remark{Remark}{
Let us note that the factor $E/(2V)$ in Eq. (\ref{eq:serlev3}) should be interpreted as the appropriate limit, i.e.
$\lim_{V\rightarrow \infty} \frac{E(V)}{2V}$, which we skip for brevity.}

\remark{Remark}{\label{rem:silnie}
Eqs. (\ref{eq:serlev1}, \ref{eq:serlev2}, \ref{eq:serlev3}) could rise the question about the presence of the factorials: in the case of the partition function factorials are implicit in the definition of $g(r)$ while the expansions of free energies contain the explicit factorials. We decided to include factorial into coefficients $g(r)$, because in such a way the variables have the natural, physical, interpretation as the number of states i.e. the number of ways in which $r$-th level of energy could be achieved. However, with the expansion of the free energies we decided to leave factorial, because it simplifies the formulas with Bell polynomials (see sec. \ref{sec:bell}). For the clarity we also present {\it explicite} factor $2$ in the definitions of $\mathfrak{g}_V^\mathcal{G}$ and $g_V^\mathcal{G}$. This factor counts the spin configurations, which are symmetrical due to the change of the every spin's sign. Factor $2$ vanishes in the limit for $V\rightarrow\infty$, which is the reason of its lack in the expansion of $\zeta_\mathcal{G}$.  }

As we mentioned in remark \ref{rem:silnie}, with this paper we obtain the exact formulas for the number of states $g_\mathcal{G}$, which have the clear combinatorial and physical interpretation. While the physical meaning is obvious, the combinatorial nature of $g_\mathcal{G}$ is stated in Facts \ref{theorem:nisko} and  \ref{theorem:wysoko} and further discussed in sec. \ref{sec:animals}.

\section{Tools and methods}\label{sec:toolsandmethods}


In the beginning of this section, firstly we introduce the most important results of the combinatorial aspects of the series expansions of the partition function and the free energy of the Ising model (partially based on chapter 13.4 in \cite{pathria} and \cite{Cipra1987}). Then we introduce the necessary for our further consideration combinatorial techniques i.e. Bell polynomials and hypergeometric functions,  and briefly discuss their properties. Next, we present the essential in our work Bell polynomials' approach introduced in  \cite{AFPF}. As an example, summarizing our previous work \cite{GSAFPF}, we consider the square lattice Ising model, in which we obtained the exact formulas for the number of states as well as combinatorial interpretation in the terms of the ideal gas of clusters \cite{AF1,AF2,GS}. We also present the generalization of the  square, triangular and hexagonal lattices --  Utiyama chequered graphs \cite{Utiyama1951}, which are considered later in the paper. The section ends with a short summarize of the most important for our work integer sequences.

In our further considerations, for brevity, we use the Iverson notation  \cite{Iverson1962} in the form introduced in \cite{Graham1994}, which is defined for a condition $P$ as follows
\begin{equation}\label{eq:Iverson}
[P]=\begin{cases}1\;\;\mathrm{for}\;\;P=\mathrm{TRUE}\\
0\;\;\mathrm{for}\;\;P=\mathrm{FALSE}
\end{cases}.
\end{equation}
Iverson notation allows us to simplify complex sums which is very helpful.

\subsection{Lattice animals and series expansions}\label{sec:animalsandseries}
Firstly, we introduce the two most typical ways to expand free energy and the partition function into a power series.
They are usually called low- and high-temperature expansions and are introduced in Definition \ref{def:lowhigh}. Secondly, we discuss combinatorial  meaning of those expansions and their connections to the problems of counting so-called lattice animals.

\subsubsection{Low- and high-temperature variables}
\label{sec:variables}
\definition{Definition}{\label{def:lowhigh}
By {\bf low-temperature expansions} we mean expansions in the variable $x$ defined as follows
	\begin{equation}\label{eq:nisko}
	x=e^{-2\beta J}.
	\end{equation}
	By {\bf high-temperature expansions} we mean expansions in the variable $v$ 
	\begin{equation}\label{eq:wysoko}
	v=\tanh (\beta J).
	\end{equation}}
Names of the both variables introduced in Definition  \ref{def:lowhigh} are consequences of the values of the temperature in which the series converges. Since $\beta\propto 1/T$ one can see that 
	\begin{equation*}
\lim_{T\rightarrow 0}x(T)=\lim_{T\rightarrow 0}e^{-2J /(k_B T)}=0,
	\end{equation*}
which means that for low temperatures $x$ is near $0$. On the other hand
	\begin{equation*}
	\lim_{T\rightarrow \infty}v(T)=\lim_{T\rightarrow \infty}\tanh\left[J /(k_B T)\right]=0,
	\end{equation*}
which is a justification of the ,,high-temperature`` label for the variable $v$.
\remark{Remark}{
Let us note that the introduced variables' names are not unique, i.e. there are more low- and high-temperature variables known in the literature \cite{domb1,domb2}. There are, for example,  high temperature star cluster expansion  \cite{Sykes1974} and, more interesting from our point of view, the expansion in the variable $\kappa$ 
\begin{equation*}
\kappa=\frac{ \sinh( 2\beta J)}{\cosh^2(2\beta J)},
\end{equation*}
which results in a beautiful expansion of the square-lattice Ising free energy 
\begin{equation}\label{eq:Viswanathan}
-\beta \varphi_\square = \ln\left[2 \cosh(2\beta J) \right] - \kappa^2\,
{}_4F_3 [1,1,\tfrac{3}{2},\tfrac{3}{2};\ 2,2,2 ;\ 16 \kappa^2 ],
\end{equation}
where ${}_4F_3$ is the generalized hypergeometric function (for definition see sec. \ref{sec:comb}). Expansion in the form given in Eq. (\ref{eq:Viswanathan}) was firstly claimed in the seminal Onsager's work (see Eq. (109c) in \cite{Onsager1944}), but, obviously, he did not use the hypergeometric function explicitly. Recently this expansion was re-discovered by Viswanathan \cite{Viswanathan2015}, who used the combinatorial properties of the hypergeometric function.
In conclusion, let us emphasise, that despite  the richness and beauty of differ variables we focus on the pair $x$ and $v$, because they are dual in the sense of Kramers-Wannier, see Fig. \ref{fig:duality}.}
Let us note that in the introduced pair $x$ and $v$, the more important for us is the low-temperature variable $x$. We introduced the high-temperature variable for the completeness of our discussion (and lattice animals interpretation through the Kramers-Wannier duality, see sec. \ref{sec:animals}), however our main goal in the paper is the investigation of the low-temperature expansions. For the applications of the high-temperature expansion for the planar Ising models see \cite{Bianca2016,domb2}.

The typical situation in the behavior of the considered lattice models (in the infinite size limit and bulk case) is the presence of the phase transition at temperature $T_c^\mathcal{G}$, which corresponds to the  variables $x_c^\mathcal{G}=e^{-2J/(k_BT_c^\mathcal{G})}$ and $v_c^\mathcal{G}=\cosh\left[J/(k_BT_c^\mathcal{G})\right]$.  They are connected to each other with the relation $x_c^\mathcal{G}=v_c^\mathcal{G}$ which is, simultaneously, the boundary between the convergence disks  of low- and high temperature expansions. The above implies 
\begin{equation}\label{eq:limit}
\lim_{n\rightarrow\infty} \frac{|a_\mathcal{G}(n)|}{n!}(x_c^\mathcal{G})^n=1.
\end{equation}
This relation may also be validated by the analytical argument - due to the Lee-Yang Theorem \cite{leeyang} one knows that the phase transition (in the temperature $T_c^\mathcal{G}$) implies the non-analytical behavior i.e. low-temperature (from the other side  high-temperature) series expansion does not converge for $x_c^\mathcal{G}=v_c^\mathcal{G}$. Below we list values of the critical low-temperature variables for the graphs used further in the article (see Fig. \ref{fig:lattice})
\begin{align}
x_c^\square&=v_c^\square=\sqrt{2}-1,\label{eq:xcS}\\
x_c^\triangle&=v_c^{\hexagon}=\frac{1}{\sqrt{3}},\label{eq:xcT}\\
x_c^{\hexagon}&=v_c^{\triangle}=2-\sqrt{3}.\label{eq:xcH}
\end{align}

\subsubsection{Low-temperature expansion}\label{sec:low}
\begin{figure}[ht!]
\begin{center}
    \includegraphics[width=0.28\textwidth]{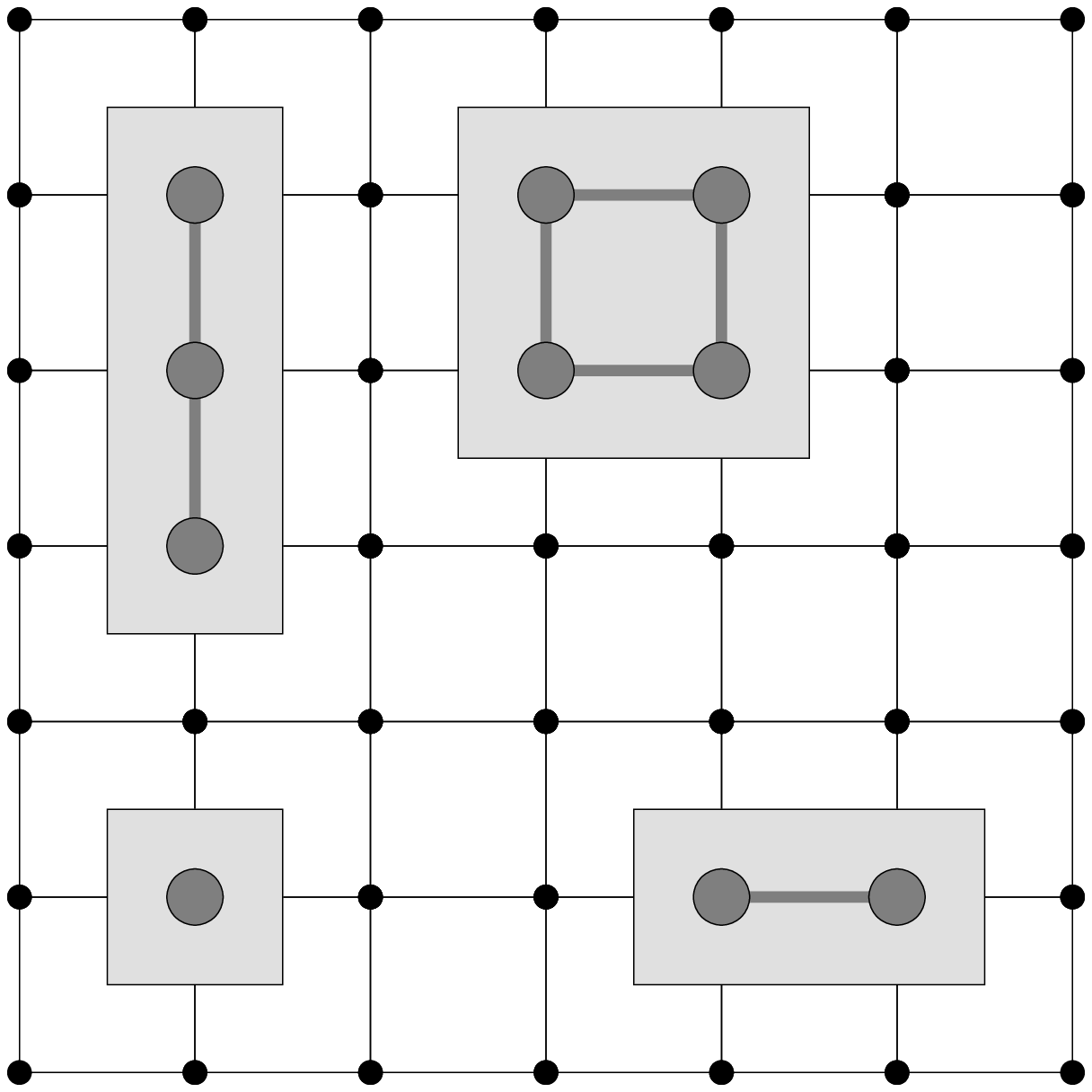}    \includegraphics[width=0.32\textwidth]{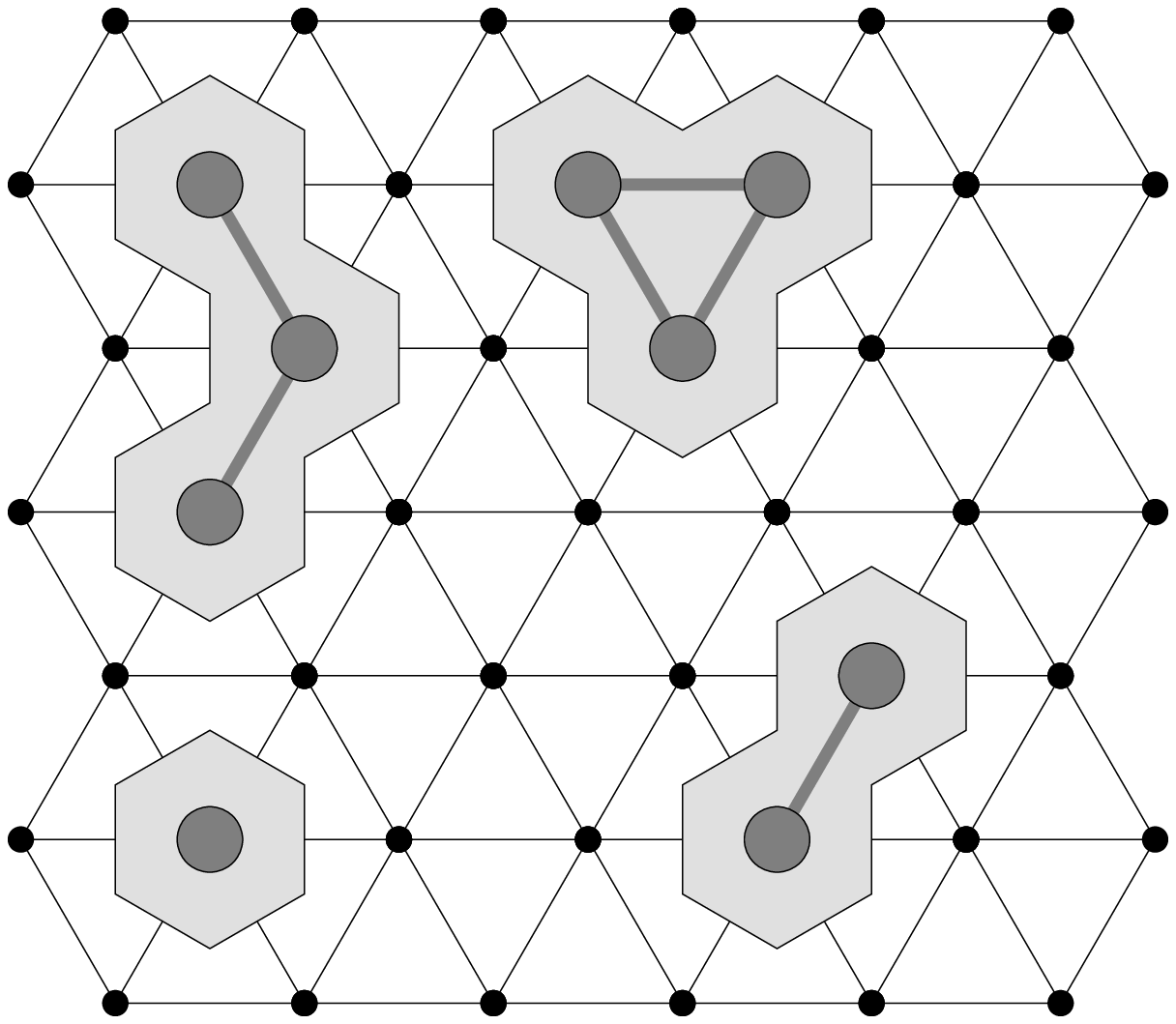}    \includegraphics[width=0.31\textwidth]{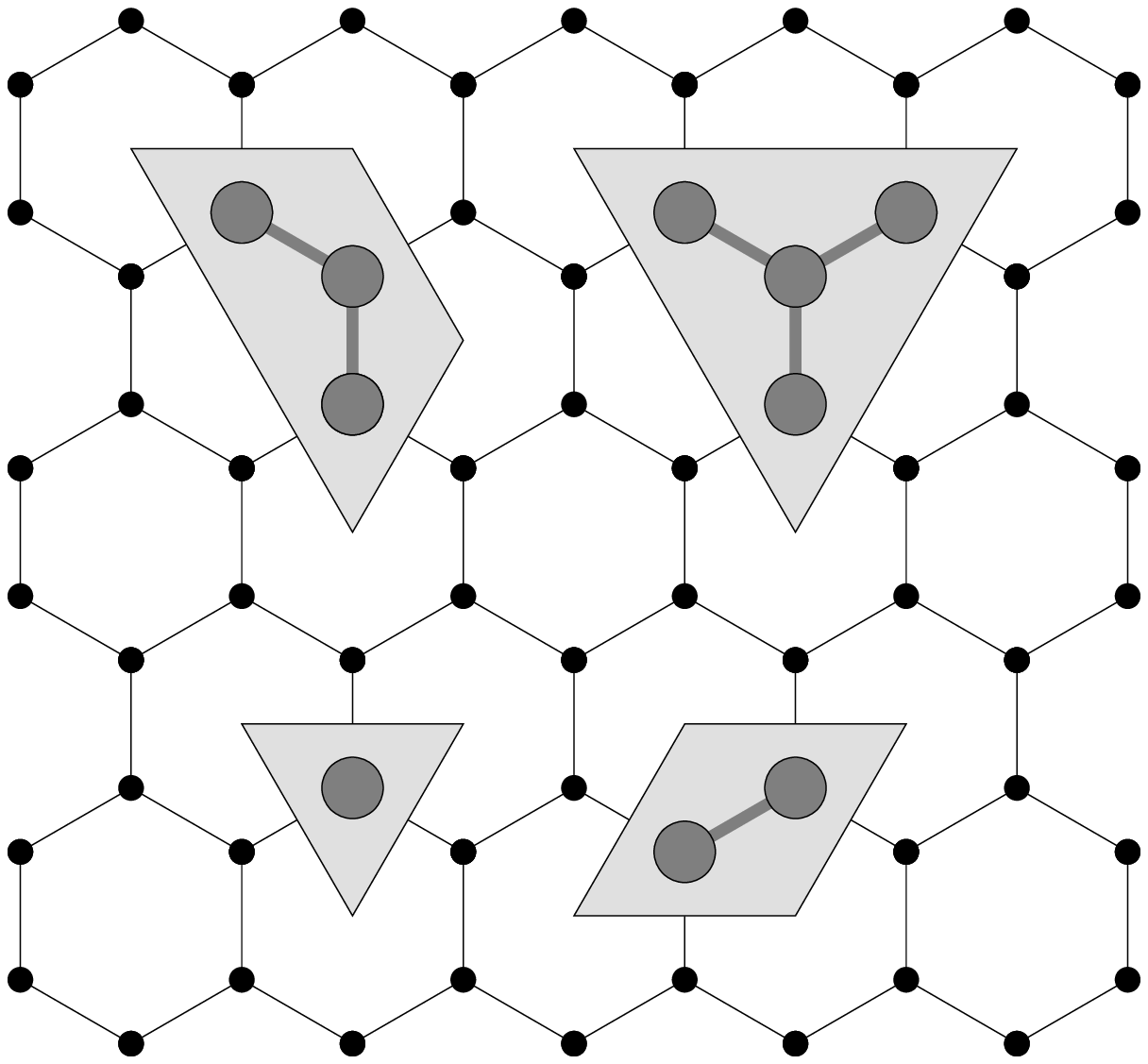}
\end{center}
	\caption{ Low-temperature animals on the lattices from Fig. \ref{fig:lattice}. }
	\label{fig:nisko}
\end{figure}

Let us start the expansion with the finite graph case and consider the Ising partition function as given by Eq.  (\ref{eq:partitionfunctionG}) 
 \begin{align*}   
 \mathfrak{Z}^{\mathcal{G}}_V&=\sum_{(s_i)}\exp\left(\beta J\sum_{\{ l,\,p\}\in E_\mathcal{G}}s_ls_p\right)=\sum_{(s_i)}e^{\beta J E}\left(\prod_{\{ l,\,p \}\in E_{\mathcal{G}}} e^{\beta J \left(s_ls_p-1\right)}\right)\stackrel{\spadesuit}{=}\\
  &\stackrel{\spadesuit}{=}e^{\beta J E}\sum_{(s_i)}\left(\prod_{\{ l,\,p \}\in E_{\mathcal{G}}} x^{ \left(1-s_ls_p\right)/2}\right)=x^{-E/2}\sum_{(s_i)}x^{\left|\{\{ l,\,p\}\in E_{\mathcal{G}}\,: s_ls_p=-1\}\right|}=\\
  =&x^{-E/2}\sum_{r=0}^E \left(\sum_{(s_i)}\Big[\big|\{\{ l,\,p\}\in E_{\mathcal{G}}\,: s_ls_p=-1\}\big|=r\Big] \right)x^r,
  \end{align*}
where for brevity we left only $(s_i)$ in the sum limit remembering that summation is taken over every configuration $(s_i)\in\{-1,\,1\}^V$. Furthermore, in $\spadesuit$ we exclude the term $e^{\beta J}$ and then substitute low-temperature variable $x$ (\ref{eq:nisko}), which, combined with the Iverson notation (\ref{eq:Iverson}), finally leads us to the low-temperature expansion
\remark{Remark: low-temperature expansion}{\label{theorem:nisko}
\begin{equation}\label{eq:lowseries}
\mathfrak{Z}_V^{\mathcal{G}}(x)=x^{- E/2}\sum_{r=0}^{E}\mathfrak{g}_V^{\mathcal{G}}(r) x^r,
\end{equation}
where coefficients $\mathfrak{g}_V^{\mathcal{G}}(r)$ describe the number of states
\begin{align*}
\mathfrak{g}_V^{\mathcal{G}}(r)=&\Big|\left\{(s_i)\,:\;\left|\big\{\{ l,\,p\}\in E_{\mathcal{G}}\,: s_ls_p=-1\big\}\Big|=r\right\}\right|,
\end{align*}
and simultaneously count the number of the site animals (see \cite{Rechnitzer2000}) with exact $r$ free bonds in the graph  $\mathcal{G}$ (see Fig. \ref{fig:nisko}).	}

\subsubsection{High-temperature expansion}\label{sec:high}
Let us recall the general identity 
\begin{equation}\label{eq:exp1}
e^{K\sigma}=\cosh K\left(1+\sigma\tanh K\right),
\end{equation}
for every $K\in\mathbb{R}$ and $\sigma=\pm 1$, which with the substitutions $K=\beta J$ and $v=\tanh(\beta J)$ (\ref{eq:wysoko}) allows one to expand the partition function (\ref{eq:partitionfunctionG}) as follows
\begin{align*}
\mathfrak{Z}_V^{\mathcal{G}}&=\sum_{(s_i)}\left(\prod_{\{ l,\,p \}\in E_{\mathcal{G}}} e^{\beta J s_ls_p}\right)=\sum_{(s_i)}\left[\prod_{\{ l,\,p \}\in E_{\mathcal{G}}} \cosh(\beta J)(1+s_ls_pv)\right]=\\
&=\left[\cosh(\beta J)\right]^{E}\sum_{(s_i)}\left[\prod_{\{ l,\,p \}\in E_{\mathcal{G}}}(1+s_ls_pv)\right].
\end{align*}
Let us note that because of the fact that $s_i^2=1$  for every $i=1,\,\dots,\,V$  one can expand the terms in the following way
\begin{align}
\mathfrak{Z}_V^{\mathcal{G}}=\left[\cosh(\beta J)\right]^{E}&\sum_{(s_i)}\Big[ P(v)+s_1P_1(v,\,s_2,\,s_3,\,\dots,\,s_{V})+\label{eq:highpoly}\\
+&s_2P_2(v,\,s_3,\,\dots,\,s_{V})+\dots+s_{V-1}P_{V-1}(v,s_{V})+s_{V}P_{V}(v)\Big].\nonumber
\end{align}
Every summand of the form $s_kP_k$ in  Eq. (\ref{eq:highpoly}) vanishes during summation over possible configurations $(s_i)$ because
\begin{align}\nonumber
&\sum_{(s_i)}s_kP_k(v,\,s_{k+1},\,\dots,\,s_{V})=\left(\sum_{s_k=\pm 1}s_k\right)\sum_{(s_i)_{i\neq k}}P_k(v,\,s_{k+1},\,\dots,\,s_{V})=\\
&=\sum_{(s_i)_{i\neq k}}P_k(v,\,s_{k+1},\,\dots,\,s_{V})-\sum_{(s_i)_{i\neq k}}P_k(v,\,s_k,\,s_{k+1},\,\dots,\,s_{V})=0.\label{eq:zerowanie}
\end{align}
Thus, one can rewrite Eq. (\ref{eq:highpoly}) due to the fact that the sumation is trivial $\sum_{(s_i)}P(v)=2^VP(v)$ and $\cosh (\beta J) = 1/\sqrt{1-v^2}$, which finally leads to
\begin{align*}
\mathfrak{Z}_V^{\mathcal{G}}(v)=\frac{2^{V}}{(1-v^2)^{E/2}} P(v).
\end{align*}
Let us note that $P(v)$ is a polynomial with degree at most $E$, which means that the final high-temperature expansion takes the following form
\begin{align*}
\mathfrak{Z}_V^{\mathcal{G}}=\frac{2^{V}}{(1-v^2)^{E/2}}\sum_{r=0}^{E}\mathfrak{q}_V^{\mathcal{G}}(r)v^r,
\end{align*}
where the very first coefficients $\mathfrak{q}_V^{\mathcal{G}}(r)$  are equal to
\begin{align}\nonumber
\mathfrak{q}_V^{\mathcal{G}}(0)&=1,\;\;\;\;\mathfrak{q}_V^{\mathcal{G}}(1)=0\;\;\;\;\mathfrak{q}_V^{\mathcal{G}}(2)=0\\
\mathfrak{q}_V^{\mathcal{G}}(r)&=\Big| \big\{ \{l_1,\,l_2,\,\dots,\,l_r\} : \{l_i,\,l_{i+1}\}\in E_\mathcal{G},\,i=1,\,\dots,\,r \big\} \Big|,\label{eq:q}
\end{align} 
where we assume that there is no self-loops i.e. $l\neq k$ for every $\{ l,\,k\}\in E_\mathcal{G}$ (which implies $\mathfrak{q}_V^{\mathcal{G}}(1)=0$) and there is at most one link between two different vertices (which implies $\mathfrak{q}_V^{\mathcal{G}}(2)=0$).  Eq. (\ref{eq:q}) can be expressed more precisely, when we introduce the following definition of self avoiding polygons (bond animals, see \cite{Rechnitzer2000}):
\definition{Definition}{
{\bf Self avoiding polygon}  in a graph $\mathcal{G}=(V_\mathcal{G},\,E_\mathcal{G})$ is a subgraphh $\mathcal{D}=(V_\mathcal{D},\,E_\mathcal{D})$  of graph $\mathcal{G}$, for which every vertex $u\in V_\mathcal{D}$ has an even degree (in graph $\mathcal{D})$. The total number of self avoiding polygons of size $r$ we denote as $\mathcal{SAP}_\mathcal{G}(r)$.}

\remark{Fact: High-temperature expansion}{\label{theorem:wysoko}
	
	\begin{equation}\label{eq:Twysoko}
\mathfrak{Z}_V^{\mathcal{G}}=\frac{2^{V}}{(1-v^2)^{E/2}}\sum_{r=0}^{E}\mathfrak{q}_V^{\mathcal{G}}(r)v^r,\;\;\;\;\mathfrak{q}_V^\mathcal{G}(r)=\mathcal{SAP}_\mathcal{G}(r).
	\end{equation}}

\subsubsection{Lattice animals}\label{sec:animals}
\begin{figure}[hb!]
\begin{center}
    \includegraphics[width=0.28\textwidth]{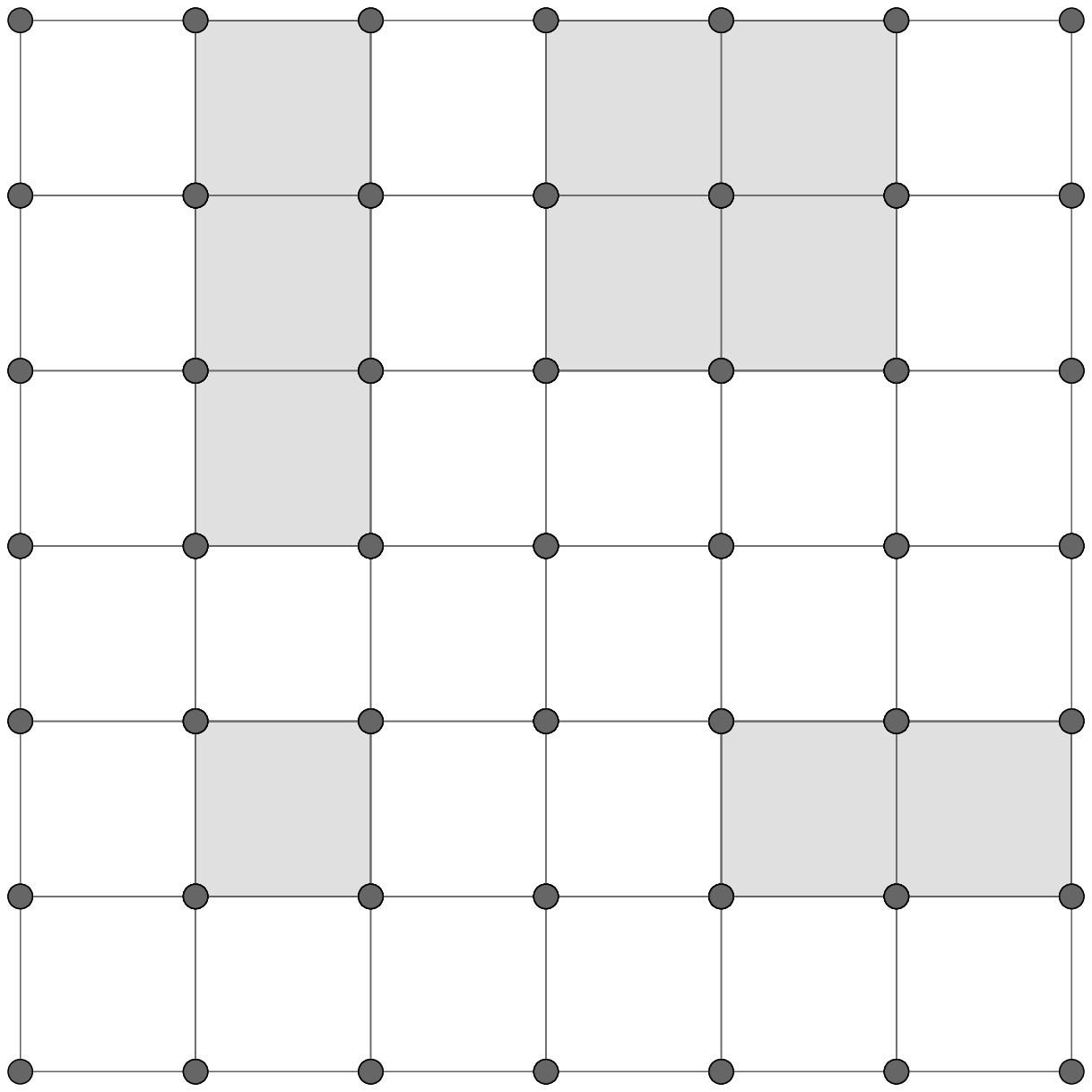}    \includegraphics[width=0.31\textwidth]{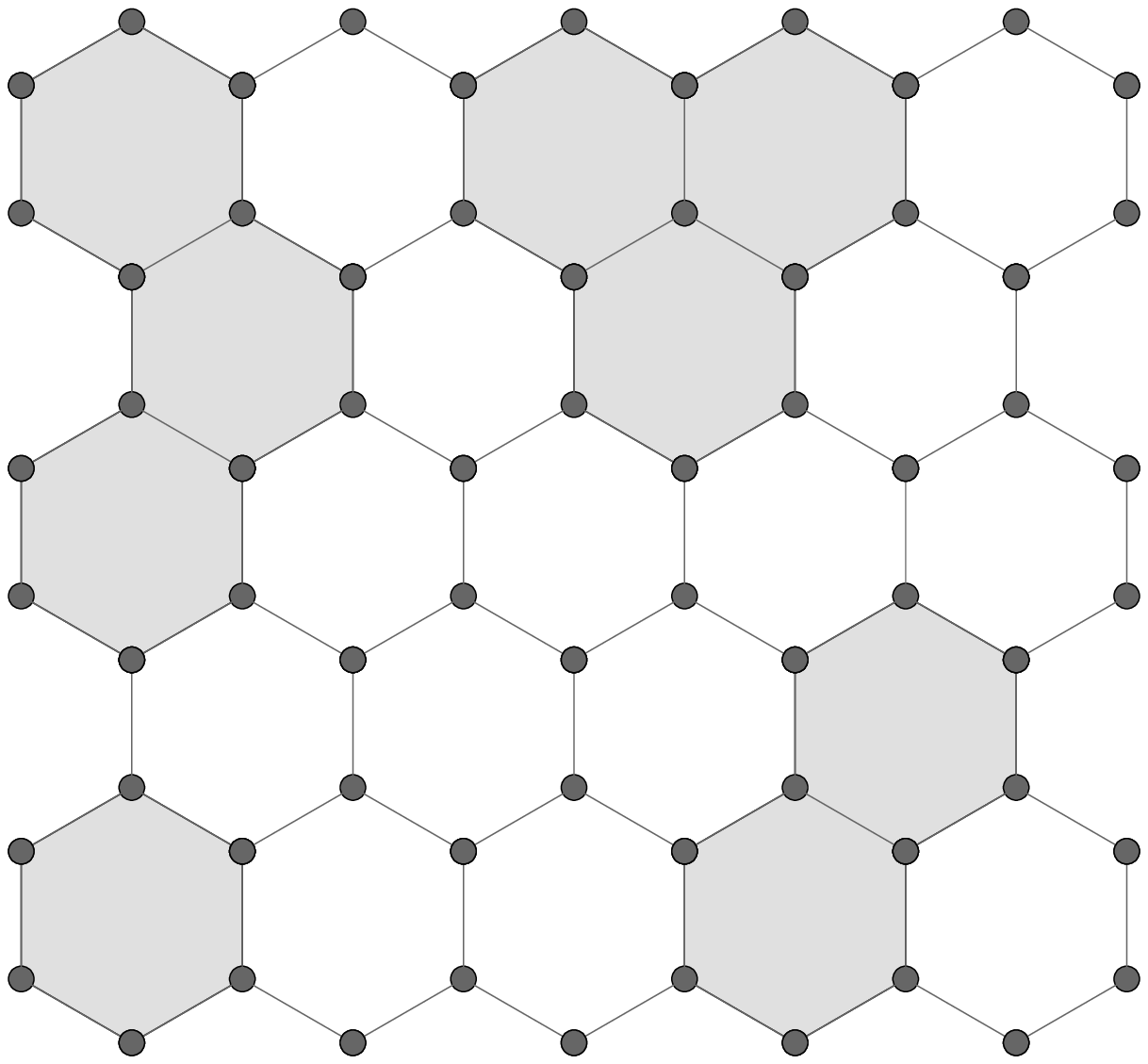}    \includegraphics[width=0.32\textwidth]{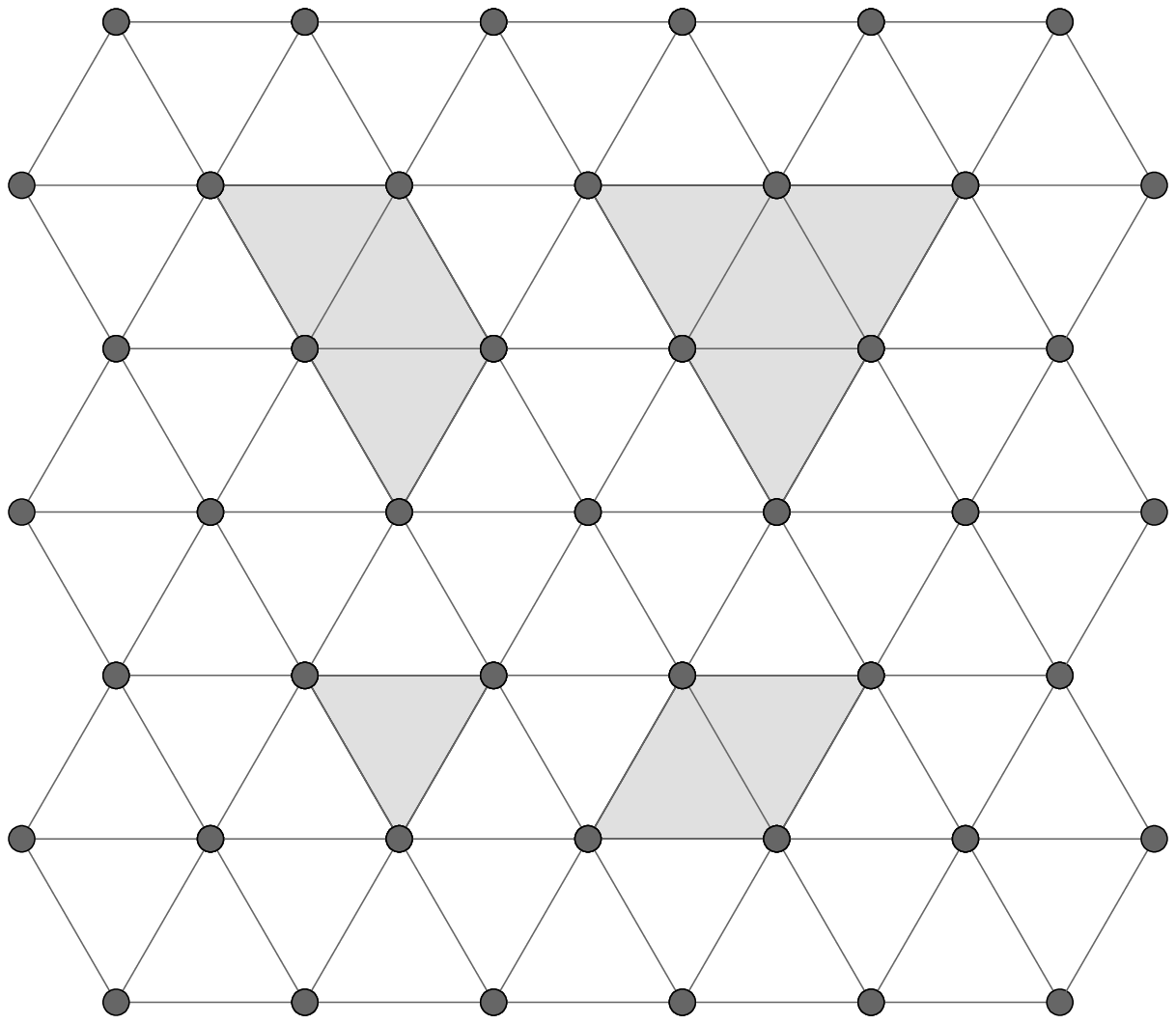}
\end{center}
	\caption{ High-temperature animals on the lattices from Fig. \ref{fig:lattice}. }
	\label{fig:wysoko}
\end{figure}

Both, the low- ($\mathfrak{g}_V$) and high-temperature series expansion's coefficients ($\mathfrak{q}_V$) has very natural combinatorial interpretation as corresponding (site- or bond-) lattice animals (for a wide discussion of the topic of lattice animals see \cite{Rechnitzer2000}). Low-temperature animals (see Fig. \ref{fig:nisko}) are the polygons being an envelope of the considered set of graph vertices. On the other hand, high temperature animals are polygons built from the elementary cells of the graphs (see Fig. \ref{fig:wysoko}). Comparing Figs. \ref{fig:nisko} and \ref{fig:wysoko} one can notice that Kramers-Wannier duality manifests with the similarity between the corresponding animals (highlighted on Fig. \ref{fig:nisko} with gray polygons).

\subsection{Combinatorial numbers, functions and polynomials}\label{sec:comb}
In the following subsection, we introduce combinatorial tools i.e. special polynomials, number and functions essential for our further considerations.

Let us start with the Bell polynomials (sometimes called also exponential and incomplete Bell Polynomials, see sec. 3.3 in \cite{comtet}), firstly introduced by Bell, see Eq.(4.5) in \cite{Bell1934}. They could be defined in two ways: analytical and combinatorial. In the first case one consider two analytic functions $f$ and $g$ (and, for simplicity, we assume $f(0)=0$) with the following expansions
\begin{equation*}
f(x)=\sum _{n=1}^{\infty }f_{n} \frac{x^{n}}{n!},\;\;\; g(x)=\sum _{n=0}^{\infty }g_{n} \frac{x^{n}}{n!}.
\end{equation*}
The question, which leads to Bell polynomials, concerns the problem of series expansion of the functions composition $(g\circ f)(\cdot)=g(f(\cdot))$
\begin{equation*}
g(f(x))=\sum _{n=0}^{\infty }b_{n} \frac{x^{n}}{n!}.
\end{equation*}
Coefficients $b_n=b_n(f_1,f_2,\dots,f_n,g_0,g_1,g_2,\dots,g_n)$ are given with the $(n,k)$-th Bell polynomials  $B_{n,k}$ with $k=1,\dots,n$ as follows
\begin{equation}\label{eq:Belldef1}
\begin{cases}
b_0=g_0,\\
b_n=\sum _{k=1}^{n}g_{k}B_{n,k}(f_{1},\dots ,f_{n-k+1})\;\;\mathrm{for}\;\;n>0,
\end{cases}
\end{equation}
which is a correct definition of the Bell polynomials, even thought it does not provide the straightforward expression for $B_{n,k}$. For that purpose let us introduce the famous Fa\'a di Bruno's formula (which was introduced years before Bell polynomials in 1855, see second equation on page 479 in  \cite{FaadiBruno1855})
\begin{equation}\label{eq:faadibruno}
\frac{d^{n}}{dx^{n}}g(f(x))=\sum_{(c_i)_n} {\frac {n!}{c_{1}!\,c_{2}!\,\cdots \,c_{n}!}}\cdot g^{(c_{1}+\cdots +c_{n})}(f(x))\cdot \prod _{j=1}^{n}\left[\frac{f^{(j)}(x)}{j!}\right]^{c_{j}},
\end{equation}
where the summation takes place over all integers $c_k\geqslant 0$, such as
\begin{equation*}
1\cdot c_{1}+2\cdot c_{2}+3\cdot c_{3}+\cdots +n\cdot c_{n}=n.\,
\end{equation*}
The composition of the functions follows Riordan's formula (for detailed historical background of the topic see \cite{Johnson2002,Craik2005})
\begin{equation}\label{eq:Riordan}
\frac{d^n}{dx^n} g(f(x)) = \sum_{k=1}^n g^{(k)}(f(x))\cdot B_{n,k}\left(f'(x),f''(x),\dots,f^{(n-k+1)}(x)\right).
\end{equation}
Combination of Eqs. (\ref{eq:Belldef1}, \ref{eq:faadibruno}, \ref{eq:Riordan}) leads to the following, combinatorial, definition of Bell polynomial
\begin{equation}\label{eq:Belldefinicja}
B_{n,k}(f_{1},f_{2},\dots ,f_{n-k+1})=\sum_{(c_i)_{n,k}} \frac{n!}{c_{1}!c_{2}!\cdots c_{n-k+1}!}\prod _{j=1}^{n}\left(\frac{f_j}{j!}\right)^{c_{j}},
\end{equation}
where the summation is taken over $n$-tuples of integers $c_i\geqslant 0$, which satisfy
\begin{align}\label{eq:c2}
\begin{cases}
 c_{1}+ c_{2}+ c_{3}+\cdots + c_{n}&=k,\\
1\cdot c_{1}+2\cdot c_{2}+3\cdot c_{3}+\cdots +n\cdot c_{n}&=n.
\end{cases}
\end{align}
Eq. (\ref{eq:Belldefinicja}) has a clear combinatorial interpretation. Conditions (\ref{eq:c2}) encrypt the problem of the partition of a set of $n$ elements into $k$ non-empty subsets (clusters). Therefore, as long as coefficients $\{f_i\}$ are non-negative (and count the number of a possible microscopic realizations of the cluster of size $i$), the Bell polynomials count the number of possible partitions of a set of $n$ elements into $k$ clusters, with $f_k$ as the number of  possible realizations of cluster of size $k$. The problem with interpretation of Bell polynomials for negative, or even complex values of $a_i$ is, however, unclear. 

Despite the definition's form (both the analytical and the combinatorial) one can prove the following Bell polynomial identity (see sec. 3.3 in \cite{comtet})
\begin{equation}\label{eq:liniowosc}
 B_{N,k}(\{c b^nf_n\})=c^kb^NB_{N,k}\left(\{f_n\}\right).
\end{equation}

The introduced polynomials $B_{n,k}$ are sometimes called incomplete, so,  for completeness' reason  let us define the complete Bell polynomials $Y_N(\{a_n\})$, which can be expressed by the incomplete Bell polynomials (see sec. 3.3 in \cite{comtet}) 
\begin{equation*}\nonumber
Y_N(\{f_n\})=\sum_{k=1}^N B_{N,k}(\{f_n\}).
\end{equation*}
Complete Bell polynomials' generating function (see \cite{Wilf1990}) is just the exponent
\begin{equation}
\exp\left[\sum_{n=1}^\infty f_n \frac{x^n}{n!} \right]=1+\sum_{N=1}^\infty Y_N(f_1,f_2,\dots),\label{eq:exponent}
\end{equation}
which is a simple consequence of Eq. (\ref{eq:Belldef1}), where $b_i=1$ for every $i$.

In the literature (see Comtet \cite{comtet}) one can find other polynomials sharing the common name ,,Bell polynomials``. The above defined are called exponential, but further in the paper we also use the so-called logarithmic Bell polynomials.  The modifications result from the change of function in the analogous of Eq. (\ref{eq:exponent}). In our work we need only logarithmic Bell polynomials $L_n$ (see Eq. (5a), p. 140 in \cite{comtet}), which, for the given analytical function $f(x)=\sum_nf_nx^n/n!$ ($f_0=1$), are defined as follows 
\begin{align}\nonumber
\ln\left(1+\sum_{n=1}^\infty f_n\frac{x^n}{n!}\right)=&\sum_{n=1}^\infty L_n\left(\{f_i\}\right)\frac{x^n}{n!}=\\\label{eq:logbell}=&\sum_{n=1}^\infty \left[ \sum_{k=1}^n(-1)^{k-1}(k-1)!B_{n,k}\left(\{f_i\}\right)\right]\frac{x^n}{n!},
\end{align}
where $B_{n,k}(\{b_i\})$ are partial Bell polynomials given by Eq.~(\ref{eq:Belldefinicja}).

Bell polynomials are a natural tool for the description of the wide range of combinatorial numbers (see Theorem B in sec. 3.2 in Comtet \cite{comtet} or detailed introduction to Bell Transform by Luschny \cite{Luschny}). As an example, let us introduce Lah numbers $L(r,k)$ (see Eq.  [3h] in Comtet \cite{comtet} or OEIS sequence \texttt{A008297}  \cite{oeis})
\begin{equation}\label{eq:Lah}
B_{r,k}(1!,2!,3!,\dots)=L(r,k)=\frac{r!}{k!}\binom{r-1}{k-1}.
\end{equation}

Apart from the Bell polynomials, in our considerations, we will also need   the generalized hypergeometric functions, which are defined through the  Pochhammer symbols (see sec. 5.2 in \cite{DLMF-gamma}) 
\begin{equation}\label{eq:pochhammer}
(a)_n=a(a+1)\dots(a+n-1)=\Gamma(a+n)/\Gamma(a),
\end{equation} 
where $\Gamma$ denotes the Euler Gamma function. 
 Given with Eq. (\ref{eq:pochhammer}) Pochhammer symbol takes the following form for negative integer arguments  (see Eq.~5.2.7 in \cite{DLMF-gamma}) 
\begin{equation}
\left(-m\right)_{n}=\begin{cases}\frac{(-1)^{n}m!}{(m-n)!},&0\leq n\leq m,\\
0,&n>m.\end{cases}\label{eq:Pochprop}
\end{equation}
Eq. (\ref{eq:Pochprop}) allows one to define  generalized hypergeometric function $\!_{p}\!F_{q}$ (see. \cite{DLMF-conhypfun})
\begin{equation}\label{eq:pFq}
_{p}\!F_{q}(a_1,\,a_2,\,\dots,\,a_p;b_1,\,b_2,\,\dots,\,b_q;z)=\sum_{k=0}^\infty\frac{(a_1)_k\cdots (a_p)_k}{(b_1)_k\cdots (b_q)_k}\frac{z^k}{k!},
\end{equation}
where $(a_i)_k$ and $(b_i)_k$  are Pochhammer symbols, introduced in Eq. (\ref{eq:pochhammer}).

\subsection{Bell polynomials' approach}\label{sec:bell}
In this work we base on the Bell polynomials' approach in the form introduced in \cite{AFPF}, which  allows one to obtain exact formulas for the number of states for the models in which one knows the series expansion of the free energy.  Which means that our first goal is to expand Eqs. (\ref{eq:phiS}, \ref{eq:phiT}, \ref{eq:phiH}). Furthermore, this technique allows to reinterpret the considered lattice models as a gas of clusters models \cite{AF1,AF2,GS}. As an illustration of the Bell polynomials' approach, we consider two-dimensional square lattice and encapsulate our previous work \cite{GSAFPF}.   However, let us note, that in the most of the notations in the current section one can replace $\square$ by $\mathcal{G}$  and treat it like a pedagogical introduction to the main part of the paper.

\subsubsection{Free energy expansion} 
 
We start with the famous Onsager formula for the bulk free energy $\varphi_\square$ given by Eq. (\ref{eq:phiS}), and then expand it into low-temperature power series (for a detailed discussion see \cite{GSAFPF})
\begin{align}\nonumber
	-\beta \varphi_\square(x)=&-\ln x+x^4+2x^6+\frac{9}{2}x^8+12x^{10}+\frac{112}{3}x^{12}+130x^{14}+\dots=\\ =& -\ln x + \sum_{n=1}^\infty a_\square(n)\frac{x^n}{n!}.\label{eq:fS}
\end{align}
It was shown in \cite{GSAFPF} that coefficients $a_\square(n)$ for $n=1,\,2,\,\dots$ are given as
\begin{align}\label{eq:San}
&a_\square(2n-1)=0,\\
\nonumber
	&a_\square(2n)=\frac{(2n)!}{2}\sum_{(c_i)}\binom{c_1+c_2+c_3+c_4}{c_1,\,c_2,\,c_3,\,c_4}\frac{(-1)^{c_2+c_3+c_4-1}2^{c_2}}{c_1+c_2+c_3+c_4}\binom{c_1+c_3}{\frac{c_1+c_3}{2}}^2,
\end{align}
where the summation is taken over all integer quadruples $c_1,c_2,c_3,c_4\geq 0$, which satisfy condition $c_1+2c_2+3c_3+4c_4=2n$. Eq. (\ref{eq:San}) gives
\begin{align*}
	\left\{\frac{a_\square(n)}{n!}\right\}=&\left\{0,\,0,\,0,\,1,\,0,\,2,\,0,\, \frac{9}{2},\,0,\,12,\,0,\,\frac{112}{3}, \,0,\,130,\,0,\,\frac{1961}{4},\dots\right\},
\end{align*}
which agrees with the first terms of expansion in the first line of Eq. (\ref{eq:fS}).

\subsubsection{Exponential formula and Bell polynomials}\label{sec:bellbell}

Let us recall, that our goal is to find the expansion of the partition function. However, we have already obtained the expansions of the free energy, which is given by Eq.~ (\ref{eq:San}). Therefore, we need to connect those two types of coefficients and such a connection is given by Eq. (\ref{eq:partitionfunctionB}) i.e.
\begin{equation}\label{eq:exp}
\zeta_\square=\exp\left[-\beta \varphi_\square \right],
\end{equation}
which  is, in fact, the generating function of the complete Bell polynomials (compare  Eq.  (\ref{eq:exponent})), thus
\begin{align}
	\zeta_\square(x)=\exp\left[-\ln x+\sum_{n=1}^\infty \frac{a_\square(n)x^n}{n!}\right]=\frac{1}{x}\left[ \sum_{r=0}^{\infty}\underbrace{\frac{1}{r!}Y_r(\{a_\square(n)\})}_{=g_\square(r)}x^r\right],\label{apf3}
\end{align}
where $Y_r(\{a_\square(n)\})$ denotes $r$-th complete Bell polynomial (see sec. \ref{sec:comb}). The main result of the work \cite{AFPF} is  the following formula for the number of states $g_\square(N)$ in terms of the coefficients of the free energy series expansion $a_\square(n)$ 
\begin{equation}\label{eq:Sg}
g_\square(N)=\frac{1}{N!}Y_N(\{a_\square(n)\}).
\end{equation}
Knowing, that the number of states is given by the Bell polynomials, let us take a closer look at their combinatorial meaning. As we mentioned in previous sec. \ref{sec:comb}, Bell polynomials have two equivalent definitions: they arise through their exponential-generating function or from the combinatorial consideration of the number of possible partitions of the set with $r$ elements into $k$ non-empty clusters. For the better explanation let us introduce the following example
\example{Example}{\label{example:bell}
Let us consider a polynomial $B_{6,2}$, which takes the form
\begin{equation}
B_{6,2}(w_{1},w_{2},w_{3},w_{4},w_{5})=6w_{1}w_{5}+15w_{2}w_{4}+10w_{3}^{2}.
\end{equation}
There are $6$ possibilities of partitions with clusters of sizes $5$ and $1$, $15=\binom{6}{4}$ partitions with blocks of sizes $2$ and $4$ and $10=\frac{1}{2}\binom{6}{3}$ with clusters of size $3$. Let us note that the clusters  as well as the set's elements are indistinguishable (as long as they have the same number of elements). The non-negative variables $w_i\geqslant 0$ counts the number of the possible microscopic realizations of the clusters of size $i$.}

\subsubsection{Perfect gas of clusters}\label{sec:pgc}

As long as free-energy coefficients $a_\square(n)$ are non-negative (compare Example \ref{example:bell}) one could see \cite{AFPF} in the Bell polynomials in Eq. (\ref{apf3}) partition function of the corresponding  perfect gas of clusters model \cite{AF1,AF2,GS}. This model is a generalization of the ideal gas concept, where particles interact only within the same cluster.  We checked in our previous work \cite{GSAFPF} that this correspondence occurs for the Ising model on the two-dimensional square lattice. In that case, the perfect gas of clusters interpretation allows one for a new insight into the nature of the paramagnetic-ferromagnetic phase transition in the terms of thermodynamic preferences for clusters. Unfortunately, this assumption does not need always be satisfied. As we see in sec. \ref{sec:triangular} for Ising model on the triangular lattice, the coefficients of the free energy $a_\triangle(n)$, may have any sign, which depends on $n$. 

Let us shed some light on the perfect gas of clusters correspondence to the square lattice Ising model. For that purpose we introduce the probability   \begin{equation}\label{eq:prob}
P_\square(N,x)=\frac{2g_\square(N)x^{N}}{\zeta_\square(x)}.
\end{equation}
which is the distribution of the system's energy (both in the Ising model and the perfect gas of clusters) and measures the probability of energy $2JN$ above the ground state.  In terms of the Bell polynomials, Eq. (\ref{eq:prob}) takes the following form
\begin{equation}\label{pn2}
P_\square(N,x)=\frac{Y_N(\{a_\square(n)x^n\})/N!}{1+\sum_{r=1}^\infty Y_r(\{a_\square(n)x^n\})/r!}.
\end{equation}

Let us note that in Bell polynomial in Eq. (\ref{pn2}) one can distinguish the thermodynamical preferences of the existence of a cluster of size $N$ i.e. the number of macroscopic realization of such clusters equals $a_\square(N)x^N/N!\simeq \left(x/x^\square_c\right)^N$ (see Eq. (\ref{eq:aS-asym}) in the next section) , which results from  Eq. (\ref{eq:limit}) for $N\gg 1$.  The above indicates that passage through the critical temperature changes the nature of the preferences: for $x<x^\square_c$ the Ising model remains in the ferromagnetic state and the preferences of clusters decrease with the size of a cluster. However for the paramagnetic case for Ising model ($x>x^\square_c$) the preferences in the perfect gas of clusters grows with $N$.  The phase transition occurs at the temperature for which the preferences do not depend on clusters' size.

\subsubsection{Asymptotics}\label{sec:asympS}
Eq. (\ref{pn2}) is a general formula for both models: Ising and perfect gas of clusters with two different interpretations, but for now let us focus on the perfect gas of clusters case. The coefficients $\{a_\square(n)x^n\}$ divided by $n!$ (because of the indistinguishability of the energy portions) may be interpreted as a thermodynamic preferences for the existence of clusters of a given size.  Those coefficients (for even $n$) follow  the asymptotic relation from Eq. (\ref{eq:limit}) where $x^\square_c=\sqrt{2}-1$ (see Eq. (\ref{eq:xcS})) is the low-temperature variable  for the critical temperature (see sec. \ref{sec:variables}). The above implies that (for even $n$) the thermodynamic preferences asymptotically follow the rule
\begin{equation}\label{eq:aS-asym}
 \frac{a_\square (n)}{n!}\simeq \left(\frac{1}{x^\square_c}\right)^{n}=\frac{1}{\left(\sqrt{2}-1\right)^n}=\left(\sqrt{2}+1\right)^n,
\end{equation}
which, with Eq. (\ref{eq:Sg}) lead to the asymptotic form of the number of states\footnote{Let us note the misprint in our previous work \cite{GSAFPF}, where in Eq.  (21) we missed $1/2$ factor.}  for even $N$ 
\begin{align}
g_\square(N)\simeq&\frac{1}{N!}\sum_{k=1}^N B_{N,k}\left(0,\,\frac{2!}{(x_c^\square)^2},\,0,\,\frac{4!}{(x_c^\square)^4},\,\dots\right)\stackrel{\spadesuit}{=}\nonumber\\
\stackrel{\spadesuit}{=}&\frac{(x_c^\square)^{-N}}{N!}\sum_{k=1}^N B_{N,k}\left(0,\,2!,\,0,\,4!,\,\dots\right)\stackrel{\clubsuit}{=}\nonumber\\\stackrel{\clubsuit}{=}&\frac{(x_c^\square)^{-N}}{N!}\sum_{k=1}^N \left[\sum\limits_{\substack{c_2+c_4+\dots=k \\ 2c_2+4c_4+\dots=N}}\frac{N!}{c_2!c_4!\dots }\left(\frac{2!}{2!}\right)^{c_2}\left(\frac{4!}{4!}\right)^{c_4}\left(\frac{6!}{6!}\right)^{c_6}\dots\right]\stackrel{\diamondsuit}{=}\nonumber\\
\stackrel{\diamondsuit}{=}&\frac{(x_c^\square)^{-N}}{(N/2)!}\sum_{k=1}^{N/2} B_{N/2,k}\left(1!,\,2!,\,3!,\,4!,\,\dots\right)\stackrel{\heartsuit}{=}\frac{(x_c^\square)^{-N}}{(N/2)!}\sum_{k=1}^{N/2} L(N/2,\,k)\label{eq:Sgasymp1}
\end{align}
where in $\spadesuit$ we apply Eq. (\ref{eq:liniowosc}). Moreover in $\clubsuit$  we realize that every odd enumerative coefficient in the Bell polynomial $B_{N,k}\left(0,\,2!,\,0,\,4!,\,\dots\right)$ is equal to zero because of the fact that enumerative variables $c_i$ satisfy $c_{2l+1}=0$ for $l=1,\,2,\,\dots$. Furthermore in $\diamondsuit$ we rearrange the summation by simply dividing the last condition for $c_n$ by factor $2$. Finally, one can see Bell polynomials $B_{N/2,k}\left(1!,\,2!,\,3!,\,4!,\,\dots\right)$ which are equal to Lah numbers $L(N/2,k)$ (see Eq. (\ref{eq:Lah}) in sec.  \ref{sec:comb}). One, finally, gets
\begin{align}
g_\square(N)\simeq(x_c^\square)^{-N}\sum_{k=1}^{N/2} \frac{(N/2-1)!}{k!(k-1)!(N/2-k)!},\label{eq:Sgasymp2}
\end{align}
where we can express the summands as a Pochhammer symbols (see sec. \ref{sec:comb})
\begin{equation}
\sum_{k=1}^{N/2} \frac{(N/2-1)!}{k!(k-1)!(N/2-k)!}=\sum_{k=0}^{N/2-1} \underbrace{\frac{(N/2-1)!}{(N/2-1-k)!}}_{\stackrel{\spadesuit}{=}(-1)^k(1-N/2)_k}\underbrace{\frac{1}{(k+1)!}}_{=1/(2)_k}\frac{1}{k!},\label{eq:Sgasymp3}
\end{equation}
where in $\spadesuit$ we use their property given by Eq. (\ref{eq:Pochprop}) in Sec. \ref{sec:comb}.
Eqs. (\ref{eq:Sgasymp2},~\ref{eq:Sgasymp3}) can be further simplified, because due to Eq. (\ref{eq:Pochprop}) we can expand the summation limit to infinity. Therefore the added terms are equal to zero
\begin{align}
g_\square(N)\simeq(x_c^\square)^{-N}\sum_{k=0}^{\infty} \frac{(1-N/2)_k}{(2)_k}\frac{(-1)^k}{k!}=(x_c^\square)^{-N}\;_{1}\!F_{1}\left(1\!-\!\frac{N}{2};2;-1\right),\label{eq:Sgasymp4}
\end{align}
where $\!_{1}\!F_{1}$ is a generalized hypergeometric function  (see. Eq. (\ref{eq:pFq}) in Sec. \ref{sec:comb}). 
The same procedure leads to the asymptotic form of the thermodynamical probability (\ref{pn2})
\begin{equation}\label{eq:probS}
\mathbb{P}_\square(N,x)\simeq\begin{cases}\frac{\left(\frac{x}{x_c^\square}\right)^{\!N}\!_{1}\!F_{1}\left(1\!-\!\frac{N}{2};2;-1\right)} {1\!+\!\sum_{r=1}^\infty\left(\frac{x}{x_c^\square}\right)^{2r}\!_{1}\!F_{1}(1\!-\!r;2;-1)}\;\;&\mathrm{for\, even\, }n,\\0\;\;&\mathrm{for\, odd\, }n.\end{cases}
\end{equation}

\remark{Remark}{\label{rem:uniwersalnyciag}
	Let us note that in the previous considerations (see e.g. Eq.~(\ref{eq:Sgasymp4})) naturally appears OEIS \cite{oeis} integer sequence \texttt{A000262}    
	\begin{equation*}
\left\{\ell!_{1}\!F_{1}(1-\ell;2;-1)\right\}_{\ell=1,\,\dots}=\{1,\, 3,\, 13,\, 73,\, 501,\, 4051,\, 37633,\, 394353,\dots\}.
	\end{equation*}
This means that every (not only for the square lattice) asymptotic expansion of number of states is proportional to the number of {\it sets of lists} i.e. the number of partitions of $n$-element set into any number of ordered subsets. Surprisingly, the same sequence appears in the problems of the boson ordering, see \cite{Blasiak2006}.}

\subsubsection{Finite lattices}\label{sec:Sfin}
Let us apply the introduced approach for the description of the Ising model on the finite (square) lattice with $V=M\times M$ vertices.  We recall Eq. (\ref{eq:serlev1}) which connects the number of states $\mathfrak{g}_V^\square(N)$, free energy $\mathfrak{F}_V^\square(N)$ and the partition function $\mathfrak{Z}_V^\square(N)$ 
\begin{equation*}
\mathfrak{Z}^\square_V(x)=\exp\left[-\beta \mathfrak{F}_V^\square(x)\right]\!=\!\frac{2}{x^{V}}\sum_{r=0}^{V}\mathfrak{g}_V^\square(r)x^r,
\end{equation*}
where the free energy $\mathfrak{F}_V^\square(x)$ expands into power series according to Eq. (\ref{eq:serlev1})
\begin{equation*}
\mathfrak{F}_V^\square(x)=-V\ln x+\sum_{n=1}^\infty\frac{ \mathfrak{a}_\square(n)}{n!}x^n.     
\end{equation*}
The natural way for the approximation of the free energy on the finite lattice is its replacement by the bulk version (infinite lattice) $\varphi^\square_V\approx\varphi_\square$, which results in
\begin{equation}\label{eq:Ff}
\mathfrak{F}_V^\square(x)\approx V\varphi_\square(x),
\end{equation}
i.e. re-calculating bulk energy per-spin (see Eqs. (\ref{eq:partitionfunctionG}, \ref{eq:partitionfunctionD}, \ref{eq:partitionfunctionB}), which reproduces the free energy low-temperature coefficients
\begin{equation}\label{eq:Aa}
\mathfrak{a}_V^\square(N)\approx Va_\square (N).
\end{equation}
\remark{Remark}{\label{remark:skonczonasiec}
Let us note that for every $N<2M$ Eq. (\ref{eq:Aa}) gives the exact expression for the coefficients $\mathfrak{a}_{M^2}^\square(N)$. This is due to the fact that lattice animals of size $N<2M$  are fully included in the $M\times M$ lattice and do not {\it wrap around} the graph.}
Knowing that the Bell polynomial $Y_N$ depends only on the first $N$ arguments one realize that for  $N<M$ the number of states $\mathfrak{g}_V^\square(N)$ which is given as
\begin{align*}
\frac{1}{N!} Y_N\left(0,\,0,\,0,\,4! V,\,0,\,6!\cdot 2V,\,0,\, 8!\cdot\frac{9}{2}V,\,0,\,10!\cdot 12V,\,0,\,12!\cdot\frac{112}{3}V,\dots\right),
\end{align*}
is the exact (for that case we assume $M>12$). Furthermore, the introduced approach reproduces also partition function (consistent with the Beale's approach  for the finite size square-lattice \cite{Beale1996})
\begin{align}\nonumber
\mathfrak{Z}_V^\square(x)=\frac{2}{x^{V}}\bigg[1&+Vx^4+2Vx^6+\left(\frac{9}{2}V\!+\!\frac{1}{2}V^2\right)x^8+ (12V\!+\!2V^2)x^{10}+\\&+\left(\frac{112}{3}V\!+\!\frac{13}{2}V^2\!+\!\frac{1}{6}V^3\right)x^{12}+\dots\bigg].\label{z3}
\end{align}
Above finite-case results conclude the presentation of the Bell polynomials' approach for the square lattice case and lead us to the generalization of the considered lattices.

\subsection{Utiyama graphs}\label{sec:utiyama}
Let us introduce, following \cite{domb1}, a general class of two-dimensional lattices, which includes  square, triangular and hexagonal, i.e. Utiyama (chequered) graphs (see \cite{Utiyama1951} and \cite{Syozi1955}).

\begin{figure}[ht!]

	\centerline{\includegraphics[scale=0.7]{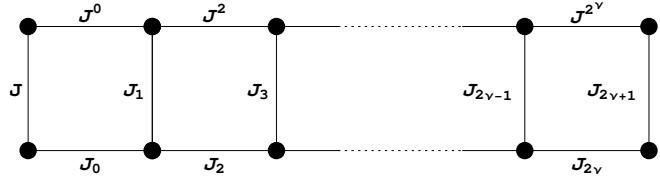}}
	
	\caption{\small Elementary Utiyama cell of size $\nu$.}
	\label{fig:szachowy}
\end{figure}

\definition{Definition}{\label{def:szachowe}
By {\bf Utiyama graphs} of size $\nu$ we mean a graph which is formed by replacing all the black squares of the chessboard by the same elementary Utiyama cell (see Fig. \ref{fig:szachowy}) of parameters 
\begin{equation}\label{eq:utidef}
J,\,J_1,\dots,\,J_{2\nu+1},\,J_0,\,J_2,\,\dots,\, J_{2\nu},\,J^2,\,\dots,\, J^{2\nu}.
\end{equation}
Coefficients $J_i$ in Eq. (\ref{eq:utidef}) can be equal to $0$, which means that the considered bond does not exist, some finite non-zero value $J$ (we assume that the lattice is isotropic, i.e. every non-zero finite $J_i$ is equal to the same value $J$)  or $\infty$, when one unite the two nodes  into one vertex.}

Ising model on Utiyama graphs could be exactly solved by the Kauffman-Onsager method \cite{domb1}. Because of the complexity of coresponding formulas let us consider the special case  $\nu=0$ (see Eq. (137)\footnote{Let us note  a misprint in the cited equation -- there is the lack of the $\prod\limits_{i:J_i=\infty}C_i$ in the denominator, which is incorrect and leads to the singularity for the triangular lattice.} in \cite{ domb1})
\begin{align}\nonumber
-\beta\varphi=& \frac{1}{2}\ln 2+\frac{1}{2(2-n_\infty)}\frac{1}{4\pi^2}\int_0^{2\pi}\int_0^{2\pi}\ln\bigg[\frac{2}{\prod\limits_{i:J_i=\infty}C_i}\bigg(1+CC_1C_0C^0+\\&+SS_1S_0S^0
-S_0S^0\cos(\theta_1+\theta_2)-SS_1\cos(\theta_1-\theta_2)+\nonumber\\&-(SS_0+S_1S^0)\cos\theta_1-(SS^0+S_1S_0)\cos\theta_2\bigg)\bigg]d\theta_1d\theta_2\label{eq:Utiyama},
\end{align}
where we use the following abbreviation
\begin{equation*}
C_i=\cosh\left(2\beta J_i\right),\;\;\;S_i=\sinh\left(2\beta J_i\right).
\end{equation*}
Furthermore $n_\infty$ is the number of bonds with $J=\infty$. One can see in Eq.~(\ref{eq:Utiyama}) a generalization of Eqs. (\ref{eq:phiS}, \ref{eq:phiT}, \ref{eq:phiH}),  with $\nu=0$ and the following values of lattice constants
	\begin{itemize}
		\item $J=\infty$ and $J_1=0$ result the square lattice,
		\item $J=\infty$ gives the triangular lattice,
		\item $J_1=0$ is hexagonal case.
	\end{itemize}
	Furthermore,  $\nu=1$ and  $J_1=\infty$ gives the kagom\'e lattice (see sec. \ref{sec:kagome}).

\subsection{Integer Sequences}
The approach presented in the article many times relies on the properties of some specific combinatorial structures and connected with them integer sequences. The OEIS (Online Encyclopedia of Integer Sequences \cite{oeis}) has an invaluable impact on our work and because of that fact, we present below the table which gathers the most important sequences occur in the paper with their OEIS numbers.

\begin{table}[ht!]
\caption{The most important integer sequences appearing in the article.}\label{tableOEIS}
\centering

 \begin{tabular}{|l|l|l|}
 \hline
 Name and notation & First terms& OEIS \\
 \hline\hline&&\\

\parbox{6cm}{Low-temperature expansion of the free energy for square lattice $a_\square(n)n!$\\ Eqs.  (\ref{eq:San}) and (8) in \cite{GSAFPF}}  & \parbox{3cm}{ $0,\,0,\,0,\,24,\,0,\,1440,$\\$0,\,181440,\,0,\,$\\$ 43545600,\,0,\,\dots$} & \texttt{A260784}\\[20pt] 
\hline&&\\
\parbox{7cm}{Low-temperature expansion of the partition function for square lattice  $g_\square(n)$\\ Eqs. (\ref{eq:Sg})cand (11) in \cite{GSAFPF}} &\parbox{3cm}{ $ 0,\,0,\, 0,\,1,\,0,\, 2,\,0,\, 5,$\\$0,\, 14,\,0,\, 44,\,0,\, 152,$\\$0,\, 566,\,0,\dots\ $} & \texttt{A002890}\\[20pt]
 \hline&&\\

Walks on the square lattice $\mathcal{S}_\square(2n)$ & \parbox{3cm}{\centering $1,\,4,\,36,\,400$\\$4900,\, 63504,$\\$853776,\,\dots$} &\texttt{A002894}\\[20pt]
 \hline&&\\

Walks on the triangular lattice $\mathcal{S}_\triangle(n)$ & \parbox{3cm}{\centering $ 1,\,0,6,\,12,\,90,\,360,$\\$2040,\,10080,\dots\ $} &\texttt{A002898}\\[20pt]
	\hline&&\\

Walks on the hexagonal lattice $\mathcal{S}_{\hexagon}(2n)$ & \parbox{3cm}{\centering $1,\,3,\,15,\,93,\,639,$\\ $4653,\,35169,\dots$\\$4900,\, 63504,$\\$853776,\,\dots$} &\texttt{A002893}\\[20pt]
		\hline

 \end{tabular}
\end{table}

\section{Applications of Bell polynomials' approach}
In this section, which is essential for our further derivations, we apply the Bell polynomials' approach to the dual (compare Figs. \ref{fig:duality}, \ref{fig:nisko} and \ref{fig:wysoko}), in the Kramers-Wannier sense, triangular and hexagonal lattices. 
Let us notice, that, while the hexagonal case is very similar to the discussed previously square lattice case (see sec. \ref{sec:bell}), the triangular lattice results in problem with the gas of cluster interpretation. However, the derived formulas work  for both cases, which is generalized to the Utiyama graphs in the next sec. \ref{sec:utiyamaapplication}.

\subsection{Triangular lattice}\label{sec:triangular}
By a triangular lattice we mean the second graph shown in Fig. \ref{fig:lattice}, which is an envelope of the (triangluar) plane tiling with Schl{\"a}fli symbol $\{3,6\}$.
\subsubsection{Free energy expansion}
The first step for applying the Bell polynomials' approach is to find the formula for the low-temperature expansion of the partition function. For that purpose, the same as in square lattice case (see sec. \ref{sec:bellbell}), we consider an integral representation of the free energy in the bulk case given by Eq. (\ref{eq:phiT}). With the substitution 
\begin{equation}\label{eq:pt}
p_\triangle=p_\triangle(\theta_1,\theta_2)=\cos\theta_1+\cos\theta_2+\cos(\theta_1+\theta_2),
\end{equation}
Eq. (\ref{eq:phiT}) transforms into the form
\begin{align}
-\beta \varphi_\triangle(x)=& \ln 2+\frac{1}{8\pi^2}\int_0^{2\pi}d\theta_1\int_0^{2\pi}d\theta_2\times\nonumber\\
&\times \ln\left[\frac{1}{8}\left(x+x^{-1}\right)^3+\frac{1}{8}\left(-x+x^{-1}\right)^3-\frac{-x+x^{-1}}{2}p_\triangle\right]=\nonumber\\
=&  \frac{1}{8\pi^2}\int_0^{2\pi}d\theta_1\int_0^{2\pi}d\theta_2 \ln\left[3x+x^{-3}+2xp_\triangle-2x^{-1}p_\triangle\right]=\nonumber\\
=&\ln x^{-\frac{3}{2}}+\frac{1}{8\pi^2} \int_{0}^{2\pi}d\theta_1\int_{0}^{2\pi}d\theta_2\ln\left[1\!-\!2p_\triangle x^2\!+(3+2p_\triangle)x^4\right].\label{eq:freenergy2}
\end{align}
Our next step is to expand the integrand function in Eq. (\ref{eq:freenergy2}) into a  series
\begin{align}
\ln&\left[1-2p_\triangle x^2\!+(3+2p_\triangle)x^4\right]\stackrel{\spadesuit}{=}\sum_{n=1}^\infty L_n\bigg(0,-4p_\triangle,0,4!(3+2p_\triangle)\bigg)\frac{x^n}{n!}\stackrel{\clubsuit}{=}\nonumber\\
\stackrel{\clubsuit}{=}&\sum_{n=1}^\infty\frac{x^n}{n!}\sum_{k=1}^n(-1)^{k-1}(k-1)!B_{n,k}\bigg(0,-4p_\triangle,0,4!(3+2p_\triangle)\bigg),\label{eq:log}
\end{align}
where in $\spadesuit$ we use  logarithmic Bell polynomials $L_n$ (see Eq. (\ref{eq:logbell})) and in $\clubsuit$ we expand them with regular Bell polynomials (see Eq. (\ref{eq:logbell}) in sec. \ref{sec:comb}). Substitution of Eq. (\ref{eq:log}) into Eq. (\ref{eq:freenergy2}) leads to
\begin{align*}
-\beta \varphi_\triangle&=\ln x^{-\frac{3}{2}}+
\sum_{n=1}^\infty\frac{x^n}{n!}\sum_{k=1}^n\frac{(-1)^{k-1}(k-1)!}{8\pi^2}\times\\&\times \int_{0}^{2\pi}d\theta_1\int_{0}^{2\pi}d\theta_2\;B_{n,k}\bigg(0,-4p_\triangle,0,4!(3+2p_\triangle)\bigg),\nonumber
\end{align*}
which combined with the expected low-temperature  expansion (see Eq. (\ref{eq:serlev3}))
\begin{align*}
-\beta \varphi_\triangle&=\ln x^{-\frac{3}{2}}+\sum_{n=1}^\infty\frac{a_\triangle(n) x^n}{n!},
\end{align*}
leads to the formula for the coefficients $a_\triangle(n)$
\begin{align}\nonumber
&\sum_{k=1}^n\frac{(-1)^{k-1}(k-1)!}{8\pi^2} \int_{0}^{2\pi}d\theta_1\int_0^{2\pi}d\theta_2\; B_{n,k}(0,-4p_\triangle,0,4!(3+2p_\triangle))\stackrel{\spadesuit}{=}\nonumber\\
\stackrel{\spadesuit}{=}&\sum_{k=1}^n\frac{(-1)^{k-1}(k-1)!}{8\pi^2} \sum_{(c_2,c4)}\frac{n!}{c_2!c_4!}\int_{0}^{2\pi}d\theta_1 \int_0^{2\pi}d\theta_2\;(-2p_\triangle)^{c_2}(3+2p_\triangle)^{c_4}, \label{eq:atriangle}
\end{align}
where in $\spadesuit$ we expand Bell polynomial $B_{n,k}$ from Eq. (\ref{eq:Belldefinicja}) and in result the summation takes  place over all pairs of non-negative integers $c_2,c_4$, which satisfy conditions 
\begin{equation}\label{eq:cond}
\begin{cases} 2c_2+4c_4=n, \\ c_2+c_4=k.\end{cases}
\end{equation}
The first conclusion from the above consideration is that for odd values coefficients vanish
\begin{equation*}
a_\triangle(2m+1)=0,\;\;m=0,\,1,\,2,\,3,\,\dots
\end{equation*}
because there is no $c_2,\,c_4$, which fulfill Eq. (\ref{eq:cond}) for odd values of $n$. One also see that for even $n$ two conditions given by Eq. (\ref{eq:cond}) may be exactly solved as
\begin{equation}\label{eq:cond2}
\begin{cases} c_2=2k-\frac{n}{2}, \\ c_4=\frac{n}{2}-k.
\end{cases}
\end{equation}
Eq. (\ref{eq:cond2}) allows one to limit the summation in Eq. (\ref{eq:atriangle}) only to $k=1,\dots,\frac{n}{2}$, because $k>\frac{n}{2}$ violates conditions (\ref{eq:cond}). Furthermore it leads to the consecutive  simplification of $a_\triangle(n)$
\begin{align}
&\sum_{k=1}^\frac{n}{2} \frac{(-1)^{k-1}(k-1)!n!}{8\pi^2(2k-\frac{n}{2})!(\frac{n}{2}-k)!}\int_{0}^{2\pi}d\theta_1 \int_0^{2\pi}d\theta_2\;(-2p_\triangle)^{2k-\frac{n}{2}}(3+2p_\triangle)^{\frac{n}{2}-k}\stackrel{\spadesuit}{=}\nonumber\\
\stackrel{\spadesuit}{=}& \sum_{k=1}^\frac{n}{2} \frac{(-1)^{k+\frac{n}{2}+1}k!n!}{8\pi^2k(2k-\frac{n}{2})!(\frac{n}{2}-k)!}\sum_{l=0}^{\frac{n}{2}-k}\binom{\frac{n}{2}-k}{l}3^{l}2^{k-l}\int_{0}^{2\pi}d\theta_1 \int_0^{2\pi}d\theta_2\;p_\triangle^{k-l}=\nonumber\\
=& \frac{n!}{2}\sum_{k=1}^\frac{n}{2} \sum_{l=0}^{\frac{n}{2}-k} \frac{(-1)^{k+\frac{n}{2}+1}3^{l}}{k}\binom{k}{\frac{n}{2}-k}\binom{\frac{n}{2}-k}{l}\underbrace{\frac{2^{k-l}}{4\pi^2} \int_{0}^{2\pi} d\theta_1 \int_0^{2\pi}d\theta_2\;p_\triangle^{k-l}}_{S_\triangle(k-l)},    \label{eq:an2}
\end{align}
where in $\spadesuit$ we expand binomial $(3+2p_\triangle)^{n/2-k}$. Let us now focus on the last  factor in Eq. (\ref{eq:an2})
\begin{align}
S_\triangle(l)=\frac{2^l}{4\pi^2}\int_{0}^{2\pi}d\theta_1\int_{0}^{2\pi}d\theta_2\left[\cos\theta_1+\cos\theta_2+\cos(\theta_1+\theta_2)\right]^l.\label{eq:rl}
\end{align}
Computation of the very first few terms of $S_\triangle(l)$ leads one  to
\begin{align*}
&S_\triangle(0)=1,\;S_\triangle(1)=0,\;S_\triangle(2)=6,\;S_\triangle(3)=12,\;S_\triangle(4)= 90,\;S_\triangle(5)= 360,\\
&S_\triangle(6)= 2040,\;S_\triangle(7)= 10080,\;S_\triangle(8)= 54810,\;S_\triangle(9)= 290640,
\end{align*}
and one can notice (for proof see \cite{observation}) that $S_\triangle(l)$ counts the number of returns to the origin in the random walk of $l$ steps on the triangular lattice (see Tab. \ref{tableOEIS}). This is well known OEIS \cite{oeis} sequence - \texttt{A002898}, and may be represented (c.f. this sequence's OEIS page) as
\begin{equation}\label{eq:rl-formula}
S_\triangle(l)= \sum_{i=0}^{l}\sum_{j=0}^i (-2)^{l-i}\binom{l}{ i} \binom{i}{j}^3.
\end{equation}
Finally,  coefficients of the free energy low-temperature series expansions are given by the following formula
\begin{align}
&a_\triangle(2n-1)=0,\nonumber\\
&a_\triangle(2n)= \frac{n!}{2}\sum_{k=1}^\frac{n}{2} \sum_{l=0}^{\frac{n}{2}-k} \frac{(-1)^{k+\frac{n}{2}+1}3^{l}}{k}\binom{k}{\frac{n}{2}-k}\binom{\frac{n}{2}-k}{l}S_\triangle(k-l).\label{eq:Tan}
\end{align}

First terms of series $a_\triangle(n)/n!$ given by Eq. (\ref{eq:Tan}) are 
\begin{align*}
	\left\{0,\; 0,\; 0,\; 0,\; 0,\; 1,\; 0,\; 0,\; 0,\; 3,\; 0,\; -\frac{3}{2},\; 0,\; 12,\; 0,\; -12,\; 0,\; \frac{181}{3},\;0,\;\;-\frac{165}{2},\dots\right\},
\end{align*}
which results with the following expansion of free energy
\begin{align}\nonumber
-\beta \varphi_\triangle=\ln x^{-\frac{3}{2}}&+x^6+3x^{10}-\frac{3}{2}x^{12}+12x^{14}+\\
&-12x^{16}-\frac{181}{3}x^{18}-\frac{165}{2}x^{20}+\dots\label{eq:aT}
\end{align}

\subsubsection{Number of states}
Number of states for the Ising model on triangular lattice has the following form (see Eq. (\ref{eq:Sg})) 
\begin{equation}\label{eq:Tg}
g_\triangle(N)=\frac{1}{N!}Y_N(\{a_\triangle(n)\}),
\end{equation}
which, combined with knowing the form of $a_\triangle$ (see (\ref{eq:Tan}) leads to 
\begin{align*}
g_\triangle=\bigg\{&0,\; 0,\; 0,\; 0,\; 0,\; 1,\; 0,\; 0,\; 0,\; 3,\; 0,\; -1,\; 0,\; 12,\; 0,\; -9,\; 0,\; 59,\; 0,\; -66,\dots\bigg\},
\end{align*}
which results in the following low-temperature expansion
\begin{align}\label{eq:series}
\zeta_\triangle=\frac{1}{x^{3/2}}\bigg(1+x^6+3x^{10}-x^{12}+12x^{14}-9x^{16}+59x^{18}-66x^{20}+\dots \bigg).
\end{align}

\subsubsection{Negative coeficients}
Negative values of the number of states given by Eq. (\ref{eq:series}) should result in negative probabilities in the system's energy distribution (see Eq. (\ref{eq:prob})), which seems ridiculous. However, despite the fact that such an idea seems to be an absurd, the concept was considered previously in the context of quantum fields theories \cite{feynman} by Feynman. To the best of our knowledge, there were no results about negative probabilities, its interpretations and possible effects in the case of lattice models, however there were some efforts made \cite{burgin} for interpretation of the events with negative probability, but adequate combinatorial interpretation, in the style of the gas of clusters, for those negative probabilities in triangular lattice case waits for its reveal. Probably  waiting for such an explanation the best way to deal with those negative coefficients is to limit probability interpretation of the number of states (for the bulk case) only for finite graphs, for which negative values do not occur (see sec. \ref{sec:Tfin}).

\subsubsection{Asymptotics}
Asymptotic behaviour of coeficients $a_\triangle$ follows for even $n$ rule similar  to Eq. (\ref{eq:limit}) ) where $x^\triangle_c$ is the low-temperature variable for the critical temperature of the model i.e. $x_c^\triangle=\frac{1}{\sqrt{3}}$ (see Eq. (\ref{eq:xcT})).
The above arguments justify asymptotic approximation of the coefficients $a_n$ in the form
\begin{equation*}
a_\triangle(n)\simeq\begin{cases}(-1)^{n/2+1}\cdot n!\cdot\left(\frac{1}{x_c^\triangle}\right)^n\;\;&\mathrm{for}\;\mathrm{even}\;n,\\0\;\;&\mathrm{for}\;\mathrm{odd}\;n,\end{cases}
\end{equation*}
which may be further simplified (see Eq. (\ref{eq:Iverson}) for the definition of used Iverson's notation) to the form
\begin{equation}\label{eq:approx}
a_\triangle(n)\simeq - [n\,\mathrm{mod}\, 2=0] n! \left(\frac{i}{x_c^\triangle}\right)^n,
\end{equation}
where $i$ is a unit imaginary number, used for brevity. Eq. (\ref{eq:approx}) is, of course, real,  because of the vanishing of coefficients for odd values of $n$.

For obtaining coefficients $g_\triangle(N)$ we proceed analogically as in the square lattice case (see Eqs. (\ref{eq:Sgasymp1}, \ref{eq:Sgasymp2}, \ref{eq:Sgasymp3}, \ref{eq:Sgasymp4}), where the only change is the formal substitution $(1/x_c^\square)^n\rightarrow -(i/x_c^\triangle)^n$ i.e.
\begin{align}
g_\triangle(N)\simeq&\begin{cases}-\frac{(-1)^{N/2}}{(x_c^\triangle)^{N}}\;_{1}\!F_{1}\left(1\!-\!\frac{N}{2};2;-1\right)\;&\mathrm{for}\;\mathrm{even}\;N,\\0\;&\mathrm{for}\;\mathrm{odd}\;N.\end{cases}
\label{eq:Tgasymp1}
\end{align}

\subsubsection{Finite lattice}\label{sec:Tfin}
\begin{figure}[hb!]
	\centerline{\includegraphics[width=0.27\textwidth]{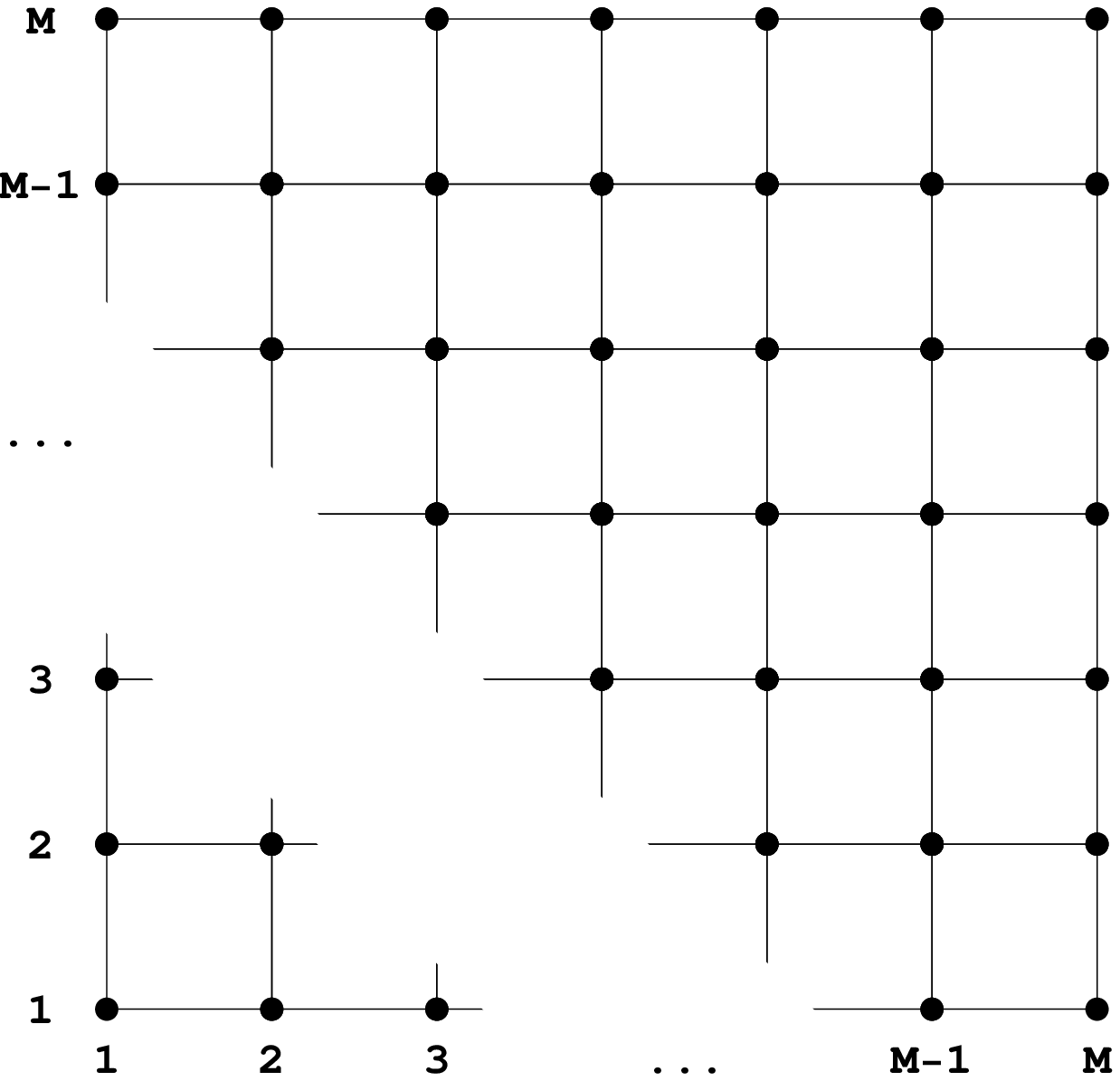}\;\;\;\;\;\;\;\;\includegraphics[width=0.27\textwidth]{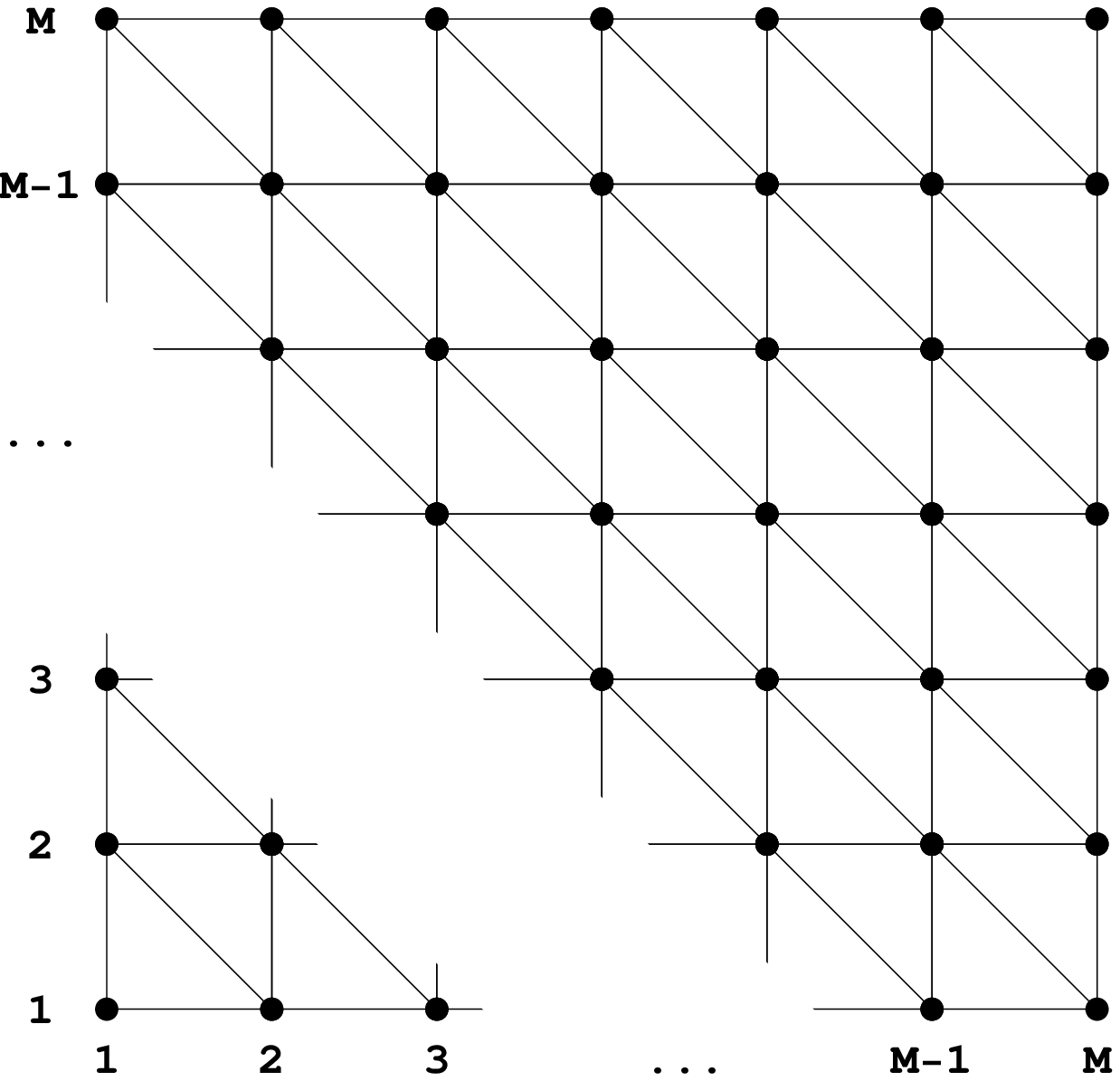}\;\includegraphics[width=0.39\textwidth]{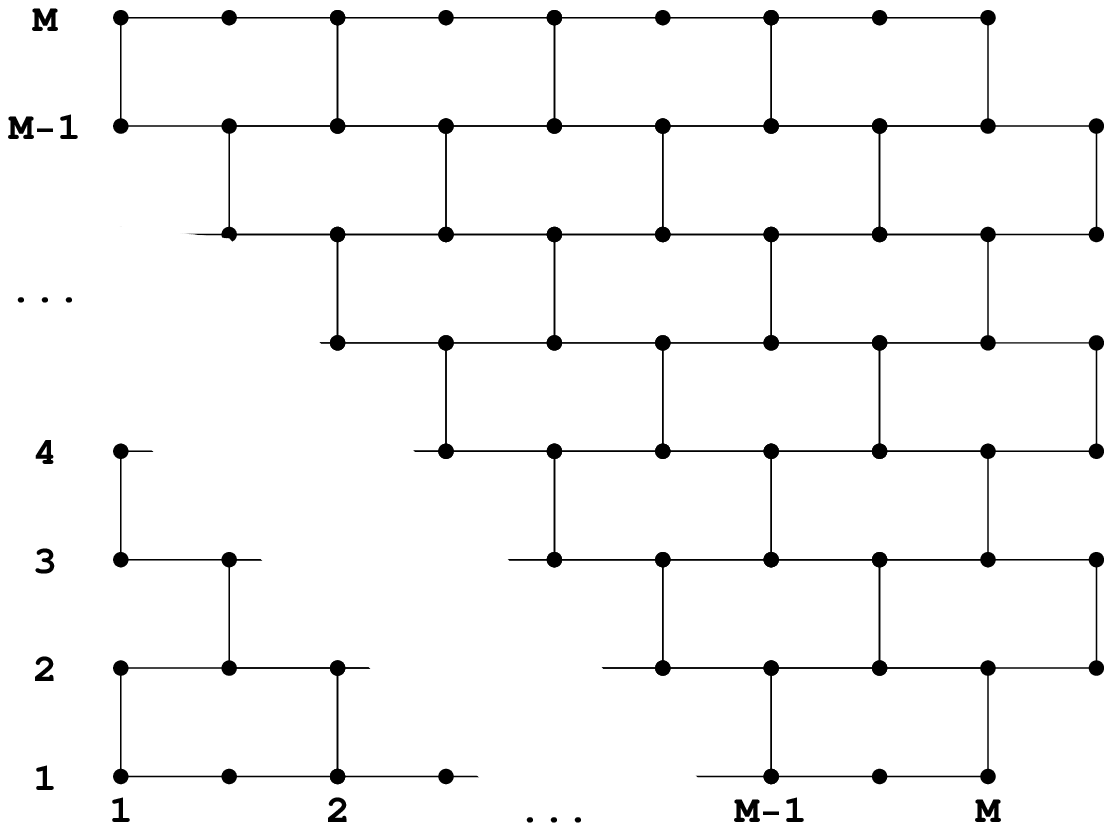}}
	\caption{\small Numbering of the vertices in the lattices from Fig. \ref{fig:lattice}. For the explanation of the different numbering rule for the hexagonal lattice see Fig. \ref{fig:N-wyjasnienie}.}
	\label{fig:N}
\end{figure}

For the finite triangular lattice of size $V=M\times M$ (see Fig. \ref{fig:N}) the partition function $\mathfrak{Z}^\triangle_V$ (see Eq. (\ref{eq:serlev1}))  has the following low-temperature expansion
\begin{equation*}
\mathfrak{Z}^\triangle_V(x)=\exp\left[-\beta \mathfrak{F}_V^\triangle(x)\right]\!=\!\frac{2}{x^{\frac{3}{2}V}}\sum_{N=0}^{V}\mathfrak{g}_V^\triangle(N)x^N,
\end{equation*}
where  $\mathfrak{F}_V^\triangle(x)$  is the system's  free energy
\begin{equation*}
\mathfrak{F}_V^\triangle(x)=-\frac{3V}{2}\ln x+\sum_{n=1}^\infty\frac{ \mathfrak{a}_\triangle(n)}{n!}x^n.     
\end{equation*}
Following the same steps as in sec. \ref{sec:Sfin} we assume that the free energy for finite graph is proportional to the bulk free energy i.e. $\mathfrak{F}_V^\triangle(x)\approx V\varphi_\triangle(x)$  which leads us to 
\begin{equation}
\mathfrak{a}_V^\triangle(n)\approx Va_\triangle (n),\label{eq:Taaprox}
\end{equation}
which (see discussion in sec. \ref{sec:Sfin}) is exact for the first terms and works  worse and worse for larger values of $n$. Eq. (\ref{eq:Taaprox}) combined with the Bell polynomials' approach leads to the approach for the finite-lattice number of states
\begin{align}
\label{z2}
\mathfrak{g}_V^\triangle(N)=\frac{1}{N!} Y_N\bigg(0,\,0,\,0,\,0,\,0,\,V\cdot 6!,\,0,\,0,\,0,\,&3V\cdot 10!,\,0,\\
& \,-\frac{3}{2}V\cdot 12!,\,0,12V\cdot 14!,\,0,\,\dots\bigg).\nonumber
\end{align}
Eq. (\ref{z2}) provides the exact formula for $N$  smaller than $4\sqrt{V}$ and it is only an approximation for higher coefficients, i.e. do not count correctly lattice animals for higher $N$. Above implies the following expansion
\begin{align}\nonumber
\mathfrak{Z}_V^\triangle(x)=\frac{2}{x^{3/2V}}\bigg[1&+Vx^6+3Vx^{10}+ \frac{1}{2}(V^2-3V)x^{12}+12V x^{14}+\dots\bigg],
\end{align}
and is adequate for sufficiently large $V$ (in that case $V\geqslant 16$).

\begin{figure}[t!]
	\centerline{\includegraphics[width=0.33\textwidth]{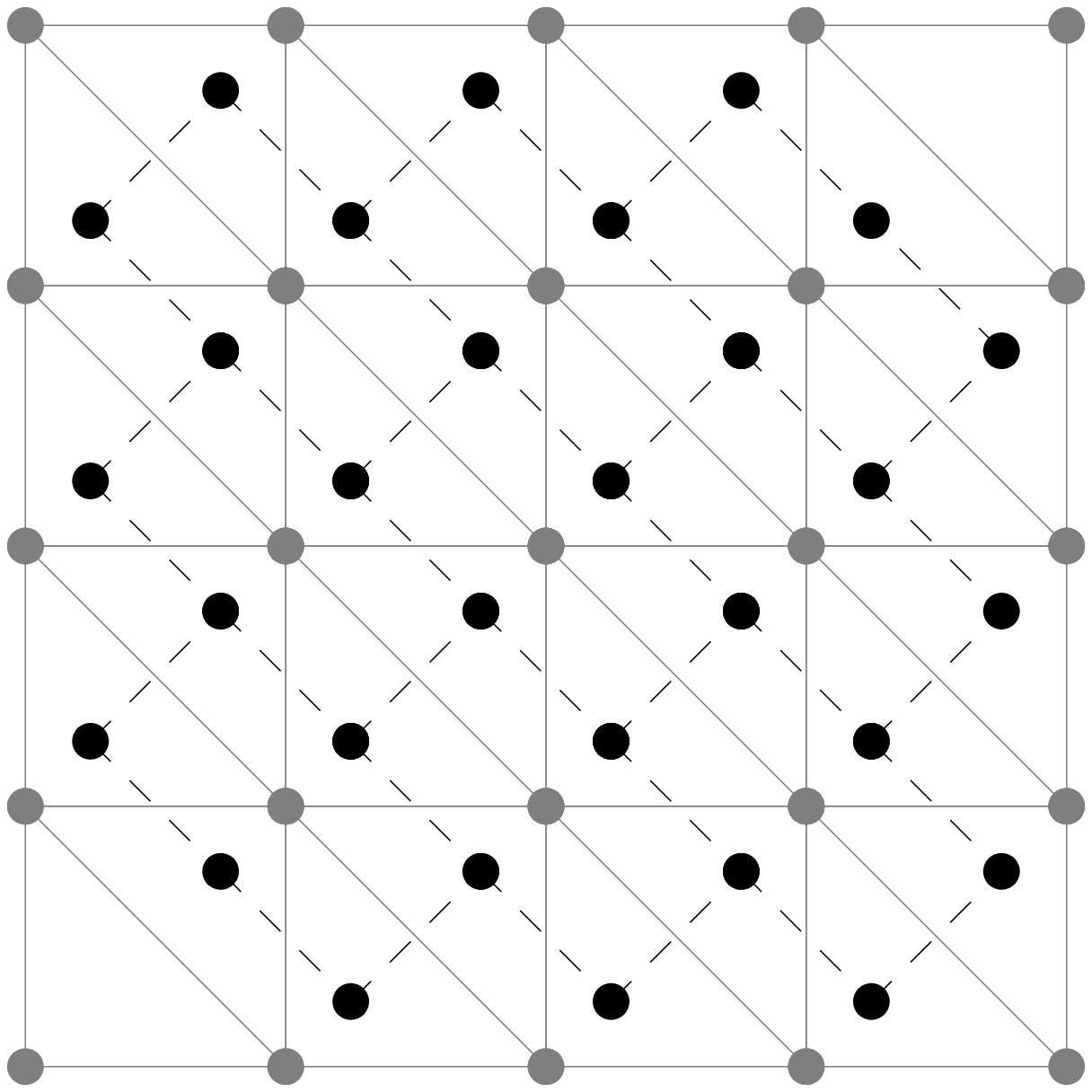}\;\;\;\;\;\;\;\;\includegraphics[width=0.33\textwidth]{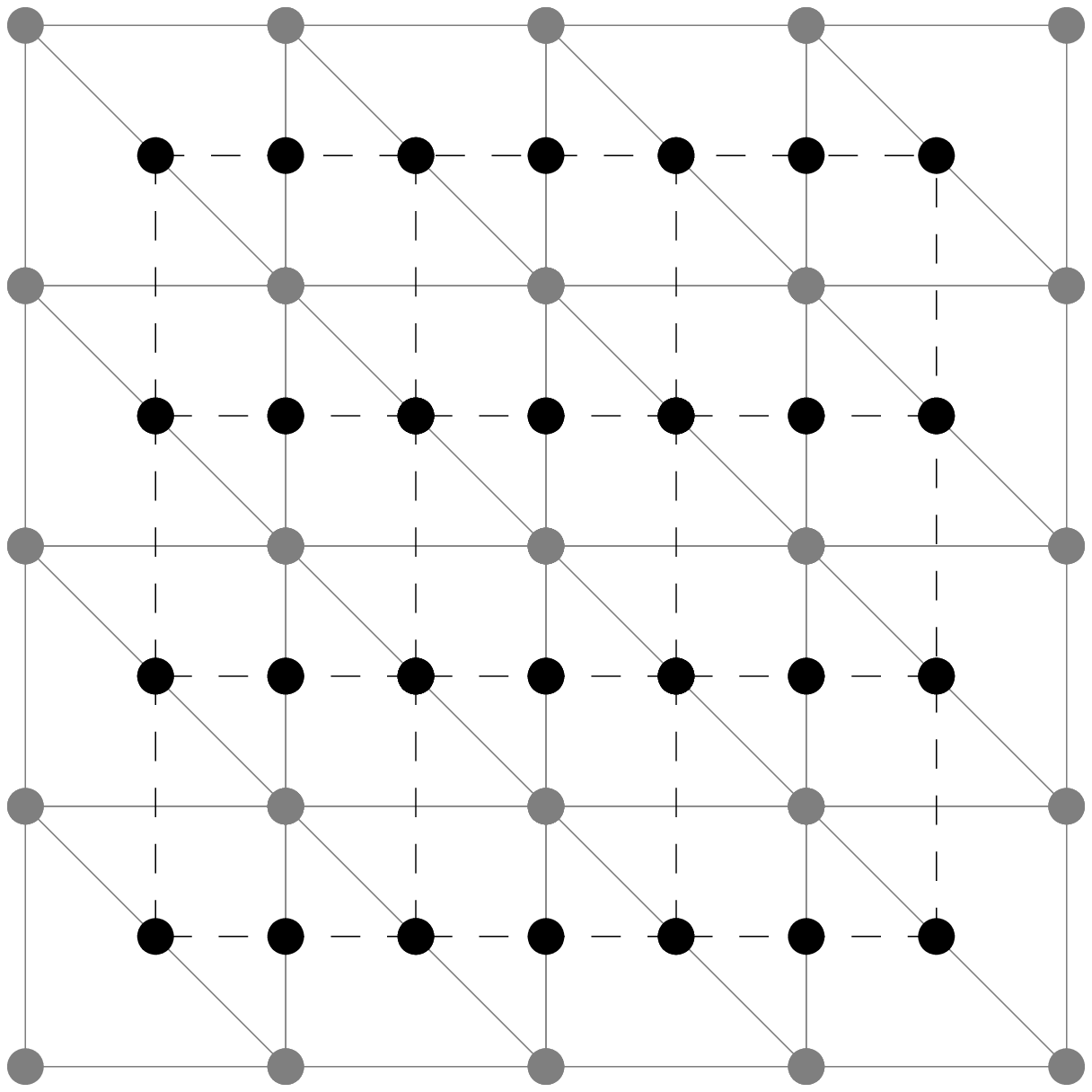}}
	\caption{\small Visualisation of the duality between lattices from Fig. \ref{fig:N}, which explains the vertices numbering rule for the hexagonal lattice. The right lattice is identical to the left one, but more clearly illustrate that horizontal dimension is effectively divided by the factor $2$.}

	\label{fig:N-wyjasnienie}
\end{figure}

\subsubsection{Lattice animals}

Let us take a closer look for coefficients $\mathfrak{g}_V^\triangle(N)$ given by Eq. (\ref{z2}) to reveal properties of counted lattice animals
\begin{align*}
&\mathfrak{g}_V^\triangle(0)=1,\, &&\mathfrak{g}_V^\triangle(1)=0,\, &&\mathfrak{g}_V^\triangle(2)=0,\, &&\mathfrak{g}_V^\triangle(3)=0,\\ &\mathfrak{g}_V^\triangle(4)=0,\, &&\mathfrak{g}_V^\triangle(5)=0,\,&&\mathfrak{g}_V^\triangle(6)=V,\, &&\mathfrak{g}_V^\triangle(7)=0,\\ &\mathfrak{g}_V^\triangle(8)=0,\, &&\mathfrak{g}_V^\triangle(9)=0,\,
 &&\mathfrak{g}_V^\triangle(10)=3V,\, &&\mathfrak{g}_V^\triangle(11)=0\\ &\mathfrak{g}_V^\triangle(12)=\frac{V^2}{2}-\frac{3V}{2},\, &&\mathfrak{g}_V^\triangle(13)=0,\,
 &&\mathfrak{g}_V^\triangle(14)=12V,\, &&\mathfrak{g}_V^\triangle(15)=0,
\end{align*}
The above polynomials in $V$ have clear combinatorial interpretation - they count the proper lattice animals (see sec. \ref{sec:animals}), i.e. the possible subsets of the vertex in the triangular lattice with $N$ unconnected (free) bonds or the $N$-polyominos (bond animals, see \cite{Rechnitzer2000}) on the dual i.e. honeycomb lattice (due to the Kramers-Wannier duality \cite{kramers}, compare Figs. \ref{fig:nisko} and \ref{fig:wysoko}). Let us examine the non-zero coefficients (for brevity we write $\mathfrak{a}_n$ for $\mathfrak{a}_V^\triangle(n)$)
\begin{align*}
\mathfrak{g}_V^\triangle (2)=&\frac{1}{2}Y_2(\{\mathfrak{a}_V^\triangle(n)\})=\frac{1}{2}\left[\mathfrak{a}_1^2V^2+\mathfrak{a}_2V\right],\\
\mathfrak{g}_V^\triangle(4)=&\frac{1}{4!}Y_4(\{\mathfrak{a}_V^\triangle(n)\})=\frac{1}{4!}\left[\mathfrak{a}_1^4 V^4+6 \mathfrak{a}_2 \mathfrak{a}_1^2 V^3+4 \mathfrak{a}_3 \mathfrak{a}_1 V^2+3 \mathfrak{a}_2^2 V^2+\mathfrak{a}_4 V\right],\\
\mathfrak{g}_V^\triangle(6)=&\frac{1}{6!}\left[\mathfrak{a}_1^6 V^6+15 \mathfrak{a}_2 \mathfrak{a}_1^4 V^5+20 \mathfrak{a}_3 \mathfrak{a}_1^3 V^4+45 \mathfrak{a}_2^2 \mathfrak{a}_1^2 V^4+15 \mathfrak{a}_4 \mathfrak{a}_1^2
   V^3+\right.\\&+\left.60 \mathfrak{a}_2 \mathfrak{a}_3 \mathfrak{a}_1 V^3+15 \mathfrak{a}_2^3 V^3+6 \mathfrak{a}_5 \mathfrak{a}_1 V^2+10 \mathfrak{a}_3^2 V^2+15 \mathfrak{a}_2 \mathfrak{a}_4
   V^2+\mathfrak{a}_6 V\right].
\end{align*}
Due to Eq. (\ref{eq:aT}) first non-zero $a_n:=a_\triangle(n)$ is $a_6=6!$ and $a_1=a_2=a_4=a_5=0$ which (with Eq. (\ref{eq:Taaprox}) i.e. $\mathfrak{a}_n=Va_n$) simplifies the above formulas to
\begin{align*}
\mathfrak{g}_V^\triangle(2)=0,\;\mathfrak{g}_V^\triangle(4)=0,\;\mathfrak{g}_V^\triangle(6)=\frac{6!V}{6!}=V.
\end{align*}
As we can infer from the above formulas -- there is no such animals which have $2$ or $4$ free nodes and there is exactly one with $6$ bonds (single vertex), so there are $V$ possibilities of putting it in lattice. In the same way one can understand $\mathfrak{g}_V^\triangle(12)$ i.e.
\begin{equation}\label{eq:g12}
\mathfrak{g}_V^\triangle(12)=\frac{1}{12!}\left(462 a_6^2 V^2+a_{12} V\right)=\frac{V^2}{2}-\frac{3V}{2},
\end{equation}
which is the reason of the negative value of $g_\triangle(12)$. We can see that the transition to the bulk case (which can be obtained formally by putting $V=1$ in Eq. (\ref{eq:g12})) results in negative values because of the negativity of $a_\triangle(12)$. 

\subsection{Hexagonal lattice} 
By a hexagonal lattice we mean the third graph from Fig. \ref{fig:lattice}, which is an envelope of the (hexagonal) plane tiling with Schl{\"a}fli symbol $\{6,3\}$.
\subsubsection{Free energy expansion}
Firstly, let us introduce variable $p_{\hexagon}$
\begin{equation}\label{eq:ph}
p_{\hexagon}=p_{\hexagon}(\theta_1,\,\theta_2)=\frac{3}{2}+\cos\theta_1+\cos\theta_2+\cos(\theta_1+\theta_2),
\end{equation}
which may seem arbitrary but it follows from the analogy to the triangular lattice case (see $p_\triangle$ given with Eq. (\ref{eq:pt}) and Eq. (\ref{eq:rl})) i.e. the expectation that the following integral
\begin{equation}\label{eq:Sh}
S_{\hexagon}(2l)=\frac{2^l}{4\pi^2}\int_{0}^{2\pi}d\theta_1\int_{0}^{2\pi}d\theta_2\;p_{\hexagon}^l,
\end{equation}
counts properly the number of walks of length $2l$ on the hexagonal lattice (see OEIS \texttt{A002893} series), which, in fact is true \cite{observation}
\begin{align*}
&S_{\hexagon}(0)=1,\;S_{\hexagon}(2)=3,\;S_{\hexagon}(4)=15,\;S_{\hexagon}(6)=93,\;S_{\hexagon}(8)= 639,\\
&S_{\hexagon}(10)= 4653,\;S_{\hexagon}(12)= 35169,\;S_{\hexagon}(14)=272835,\, \dots
\end{align*}
With the introduced notion of $p_{\hexagon}$ let us expand Eq. (\ref{eq:phiH}) with the low-temperature variable $x$
\begin{align*}
-\beta \varphi_{\hexagon}=&
\frac{3}{4}\ln 2+\frac{1}{16\pi^2}\int_0^{2\pi} d\theta_1 \int_0^{2\pi}
d\theta_2\times\\
&\ln\bigg[1+\left(\frac{x+x^{-1}}{2}\right)^3-\left(\frac{-x+x^{-1}}{2}\right)^2\left(p_{\hexagon}-\frac{3}{2}\right)\bigg]\stackrel{\spadesuit}{=}\\
\stackrel{\spadesuit}{=}&
\ln x^{-3/4}+\frac{1}{16\pi^2}\int_0^{2\pi} d\theta_1 \int_0^{2\pi}
d\theta_2\times\\
&\ln\bigg[1+(3-2p_{\hexagon})x+3x^2+(2+4p)x^3+3x^4+(3-2p)x^5+x^6\bigg]\stackrel{\clubsuit}{=}\\
\stackrel{\clubsuit}{=}&\ln x^{-3/4}+\sum_{n=1}^\infty\frac{x^n}{n!}\sum_{k=1}^n\frac{(-1)^{k-1}(k-1)!}{16\pi^2} \int_{0}^{2\pi}d\theta_1\int_{0}^{2\pi}d\theta_2\times\\
&B_{n,k}\bigg((3-2p_{\hexagon}),2!\, 3,3!(2+4p_{\hexagon}),4!\,3,5!(3-2p_{\hexagon}),6!\bigg)\stackrel{\diamondsuit}{=}\\
\stackrel{\diamondsuit}{=}&\ln x^{-3/4}+\sum_{n=1}^\infty\frac{x^n}{n!}\sum_{k=1}^n\sum_{(c_i)_{n,k}}\frac{(-1)^{k-1}(k-1)!n!}{16\pi^2c_1!c_2!c_3!c_4!c_5!c_6!}3^{c_2+c_4}2^{c_3}\times\\
&\int_{0}^{2\pi}\int_{0}^{2\pi}(3-2p_{\hexagon})^{c_1+c_5}(1+2p_{\hexagon})^{c_3}d\theta_1d\theta_2,
\end{align*}
where in $\spadesuit$ we extract fator $(2x)^{-3}$, next in $\clubsuit$ we take the advantage of the logarithmic Bell polynomials (see Eq. (\ref{eq:logbell})), and finally in $\diamondsuit$ the expand Bell polynomials from Eq. (\ref{eq:Belldefinicja}), where $\sum_{(c_i)_{n,k}}$ is a sum over integers $c_1,\,\dots,\,c_6$, which satisfies Eq. (\ref{eq:c2}), which finally allows us to write $a_{\hexagon}$ as follows
\begin{align}\nonumber
a_{\hexagon}=&\sum_{k=1}^n\sum_{(c_i)_{n,k}}\sum_{l=0}^{c_1+c_5}\sum_{r=0}^{c_3}\frac{(-1)^{k+l-1}(k-1)!n!}{16\pi^2c_1!c_2!c_3!c_4!c_5!c_6!}\binom{c_1+c_5}{l}\binom{c_3}{r}\times\\
&\times 2^{c_3}3^{c_1+c_2+c_4+c_5-l}\int_{0}^{2\pi}\int_{0}^{2\pi}(2p_{\hexagon})^{l+r}d\theta_1d\theta_2=\nonumber\\
=&\frac{n!}{2}\sum_{k=1}^n\sum_{(c_i)_{n,k}}\sum_{l=0}^{c_1+c_5}\sum_{r=0}^{c_3}\frac{(-1)^{k+l-1}(k-1)!2^{c_3-1}3^{k-c_3-c_6-l}S_{\hexagon}(2(r+l))}{c_1!c_2!c_4!c_5!c_6!(c_1+c_5)!(c_1+c_5-l)!(c_3-r)!l!r!}.\label{eq:Han}
\end{align}
First terms of series $a_{\hexagon}(n)/n!$ given by Eq. (\ref{eq:Han}) are equal to
\begin{align*}
	\left\{0,\; 0,\; 1,\; \frac{3}{2},\; 3,\; \frac{11}{2},\; 12,\; \frac{111}{4},\; \frac{208}{3},\; \frac{363}{2},\; 495,\dots\right\},
\end{align*}
which results in the following expansion of the free energy
\begin{align}\nonumber
-\beta \varphi_{\hexagon}=\ln x^{-\frac{3}{4}}+x^3+\frac{3}{2}x^{4}+3x^5+\frac{11}{2}x^6+12x^7+\frac{111}{4}x^8+\frac{208}{3}x^9+\dots
\end{align}

\subsubsection{Number of states}
Number of states for the Ising model on hexagonal lattice has the same form  as for square lattice  (see Eq. (\ref{eq:Sg})) and triangular lattice  (see Eq. (\ref{eq:Tg})) which, combined with the knowledge of the form of $a_{\hexagon}$ (see Eq. (\ref{eq:Han}) leads to 
\begin{align}\label{eq:gH}
	&g_{\hexagon}(N)=\\&\frac{1}{N!} Y_N\left(	0,\; 0,\;2\cdot 3!,\; 3\cdot 4!,\; 6\cdot 5!,\; 11\cdot 6!,\; 24\cdot 7!,\; \frac{111}{2}\cdot 8!,\; \frac{416}{3}\cdot 9!,\;\dots\right),\nonumber
	\end{align}
which results in the following expansion of the partition function
	\begin{align*}
	\zeta_{\hexagon}(x)=\frac{1}{x^{3/4}}\left(1+2x^3+3x^4+6x^5+13x^6+30x^7+72x^8+180x^9+\dots\right).
	\end{align*}
	
\subsubsection{Perfect gas of clusters}
The whole interpretation of the perfect gas of clusters described for square lattice in sec. \ref{sec:pgc} is similar for hexagonal case. The only difference is in particular values of the critical temperature $x_c^{\hexagon}$ of transition between two phases of  thermodynamic behaviour of the gas of clusters. Therefore, the cluster interpretation for the hexagonal lattice holds, unlike for the triangular lattice.

\subsubsection{Asymptotics}
Asymptotic behaviour of coefficients $a_{\hexagon}$ follows similar to the square and triangular lattices rule (\ref{eq:limit}), where $x^{\hexagon}_c$ is the low-temperature variable for the critical temperature i.e. $x_c^{\hexagon}=2-\sqrt{3}$ (see Eq.~(\ref{eq:xcH})).
The above arguments justify asymptotic approximation (see sec. \ref{sec:asympS}) of the coeficients $a_{\hexagon}(n)$ in the form
\begin{equation}\label{eq:aH-asym}
 \frac{a_{\hexagon} (n)}{n!}\simeq \left(\frac{1}{x^{\hexagon}_c}\right)^{n}=\frac{1}{\left(2-\sqrt{3}\right)^n}=\left(2+\sqrt{3}\right)^n,
\end{equation}
which leads to the following asymptotoic form of the number of states
\begin{align*}
g_{\hexagon}(N)\simeq(x_c^{\hexagon})^{-N}\;_{1}\!F_{1}\left(1\!-\!\frac{N}{2};2;-1\right),
\end{align*}
where $\!_{1}\!F_{1}$ is the generalized hypergeometric function  (see. Eq. (\ref{eq:pFq}) in Sec. \ref{sec:comb}). The same procedure leads to the asymptotic form of the thermodynamic probability (\ref{pn2})
\begin{equation}\label{eq:probH}
\mathbb{P}_\square(N,x)\simeq\frac{\left(\frac{x}{x_c^{\hexagon}}\right)^{\!N}\!_{1}\!F_{1}\left(1\!-\!\frac{N}{2};2;-1\right)} {1\!+\!\sum_{r=1}^\infty\left(\frac{x}{x_c^{\hexagon}}\right)^{2r}\!_{1}\!F_{1}(1\!-\!r;2;-1)}.
\end{equation}

\subsubsection{Finite lattice}

As we argued in sec. \ref{sec:Sfin} the asymptotic (in terms of size of the lattice $V$, not the value of energy level $N$) form of finite-lattice coefficients $\mathfrak{a}_V^{\hexagon}(n)$ (see Eq. (\ref{eq:Aa})) are equal to 
\begin{equation*}
\mathfrak{a}_V^{\hexagon}(n)\approx Va_{\hexagon} (n),
\end{equation*}
which results in the following form of the number of states $\mathfrak{g}_V^{\hexagon}(N)$ 
	\begin{align*}
	&\mathfrak{g}_V^{\hexagon}(N)=\\&\frac{1}{N!} Y_N\left(	0,\; 0,\;V\cdot 3!,\; \frac{3V}{2}\cdot 4!,\; 3V\cdot 5!,\; \frac{11V}{2}\cdot 6!,\; 12V\cdot 7!,\; \frac{111V}{4}\cdot 8!,\; \dots\right),
	\end{align*}
and finally the partition function $\mathfrak{Z}_V^{\hexagon}$ has the following  expansion
	\begin{align}\nonumber
	&\mathfrak{Z}_V^{\hexagon}(x)=\\&\frac{1}{x^{3/4V}}\left[1+V x^3 + \frac{3V}{2} x^4 + 3V x^5 + \frac{V}{2}(11+V) x^6  +3V(4+\frac{V}{2}) x^7 +\dots\right],\nonumber	
	\end{align}
which agrees with the naive calculation of the hexagonal lattice animals and ends our consideration about the dual pair of lattices: triangular--hexagonal. As we can see the analytical results hold for both cases almost in the same way as for the square lattice. However, differences in the gas of clusters interpretations occur, because of  the different combinatorics of the lattice animals on those types of lattices. The above justifies that  in the next section we apply the analytical part of the Bell polynomials' approach to the generalization of the considered lattices but without the combinatorial interpretation of the model.

\section{Formalism generalization to Utiyama graphs}\label{sec:utiyamaapplication}
Let us recall the Utiyama graphs introduced earlier in sec \ref{sec:utiyama} and consider, for brevity, the simplest case of $\nu=0$ ($\nu>0$ goes analogically), which is the case of all the  lattices from Fig. \ref{fig:lattice}.  We transform the mentioned equation (\ref{eq:Utiyama}) according to the possible values of $J_i$, which results in the following values of the hyperbolic functions for $x=\exp[-2\beta J]$

\begin{figure}[b!]
	\centerline{\includegraphics[width=0.33\textwidth]{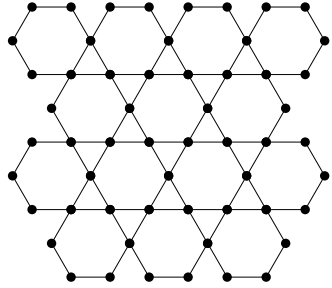}}
	\caption{\small Kagom\'e lattice -- an example of the Utiyama graph for $\nu=1$ and $J_1=\infty$.}

	\label{fig:kagome}
\end{figure}

\begin{equation}\label{eq:SC}
\begin{cases}
J_i=0 \Rightarrow &C_i=1,\; S_i=0,\\
J_i=J \Rightarrow &C_i=(x+x^{-1})/2, S_i=(x-x^{-1})/2,\\
J_i=\infty \Rightarrow &C_i=\infty,\; S_i=\infty,\; S_i/C_i=1,
\end{cases}
\end{equation}
which allows to modify Eq. (\ref{eq:Utiyama}) as follows
\begin{align}\nonumber
-\beta\varphi=&\frac{3-n_\infty}{2}\ln 2+\frac{1}{2(2-n_\infty)}\frac{1}{4\pi^2}\int_0^{2\pi}\int_0^{2\pi}d\theta_1d\theta_2\ln\Bigg[[n_\infty=0]+\nonumber\\
&+\bigg(\frac{x+x^{-1}}{2}\bigg)^{4-n_\infty}+\bigg(\frac{x-x^{-1}}{2}\bigg)^{4-n_\infty}+\nonumber\\
&-[J, J_1\neq\infty \wedge J_0, J^0\neq 0]\bigg(\frac{x-x^{-1}}{2}\bigg)^{[J_0=\hat{J}]+[J^0=\hat{J}]}\cos(\theta_1+\theta_2)+\nonumber\\
&-[J_0, J^0\neq\infty \wedge J, J_1\neq 0]\bigg(\frac{x-x^{-1}}{2}\bigg)^{[J=\hat{J}]+[J_1=\hat{J}]}\cos(\theta_1-\theta_2)+\nonumber\\
&-[J^0, J_1\neq\infty \wedge J, J_0\neq 0]\bigg(\frac{x-x^{-1}}{2}\bigg)^{[J=\hat{J}]+[J_0=\hat{J}]}\cos\theta_1+\nonumber\\
&-[J, J_0\neq\infty \wedge J_1, J^0\neq 0]\bigg(\frac{x-x^{-1}}{2}\bigg)^{[J_1=\hat{J}]+[J^0=\hat{J}]}\cos\theta_1+\nonumber\\
&-[J_0, J_1\neq\infty \wedge J, J^0\neq 0]\bigg(\frac{x-x^{-1}}{2}\bigg)^{[J=\hat{J}]+[J^0=\hat{J}]}\cos\theta_2+\nonumber\\
&+[J, J^0\neq\infty \wedge J_0, J_1\neq 0]\bigg(\frac{x-x^{-1}}{2}\bigg)^{[J_0=\hat{J}]+[J_1=\hat{J}]}\cos\theta_2\bigg)\Bigg],\label{eq:complex}
 \end{align}
which seems complex, although one can simplify the expression under the logarithm into a polynomial in variable $x$ and then apply the Bell Polynomial's procedure similar to those described in the previous sections. Because of the complexity of the Eq. (\ref{eq:complex}) in general, we illustrate that part for special case i.e. kagome\'e  lattice (see Fig. \ref{fig:kagome}). Let us note, that Eq. (\ref{eq:complex}) is the simplest case of general Utiyama formulas, it assumes that $\nu=0$ (for which examples are $\square$, $\triangle$ and ${\hexagon}$), however, the following example (kagom\'e lattice) is the special cases for $\nu=1$.

\subsection{Kagom\'e lattice}\label{sec:kagome}
The free energy of the Ising model on the kagom\'e lattice (Utiyama graph for $\nu=1$ and $J_1=\infty$, see Fig. \ref{fig:kagome}) can be represented with the following integral (see \cite{ domb1}) 
	\begin{align}
-\beta\varphi_\kappa=&\ln 2+\nonumber\\+&\frac{1}{24\pi^2}\int_0^{2\pi}\int_0^{2\pi}\ln\bigg[\frac{1}{4}\left( C^6+S^6+2C^3S^3+3C^2-2(CS^3+C^2S^2)p_\triangle\right)\bigg]\label{eq:kagome1},
	\end{align}
	where $p_\triangle$ was previously defined in Eq. (\ref{eq:pt}) and $S$ and $C$ are abbreviation for the sinh and cosh see Eq. (\ref{eq:SC}), which results in the series expansion in the following form
	\begin{align*}
	-\beta\varphi_\kappa=&\ln x^{-1}\frac{1}{24\pi^2}\int_0^{2\pi}\int_0^{2\pi}d\theta_1d\theta_2\ln\bigg[1-4 p_\triangle x^2+2(2 p_\triangle+9) x^4+\\
	&+4( p_\triangle+6)x^6+(21-4p_\triangle)x^8\bigg]=\\
	=&\ln x^{-1}+\sum_{n=1}^\infty\sum_{k=1}^n \frac{(-1)^{k-1}(k-1)!}{24\pi^2}  \int_0^{2\pi}\int_0^{2\pi}d\theta_1d\theta_2\times\\
	\times& B_{n,k}\left(0,-8p_\triangle,2\cdot4!(2p_\triangle+9),0,4\cdot 6!(p_\triangle+6),0,8!(21-4p_\triangle)\right),
	\end{align*}
	and one can see that the Bell polynomials' approach can be applied further.

\section{Summary}
We discussed the applicability of the Bell polynomials' approach for the Ising model on a wide range of planar graphs. The approach was introduced in \cite{AFPF} and previously applied e.g. to: the grand potential distribution \cite{AF1}, one-dimensional lattice gas \cite{AF2} and square-lattice Ising model \cite{GSAFPF}. With this article, we complement those results with the application to the Ising model on Utiyama graphs,  especially triangular and hexagonal lattices.  We obtained the combinatorial formulas for coefficients of the low-temperature series expansion of the free energy and partition function.  Derived results are exact for the bulk case and approximate for the finite lattices. It is worth to emphasise, that despite  the lattice topology the number of states is, approximatelly proportional to OEIS sequence A000262 (see remark \ref{rem:uniwersalnyciag} on page \pageref{rem:uniwersalnyciag}), the only difference in the asymptotic form is the value of $x_c^\mathcal{G}$.

We believe, that despite described above results our work pose a question of the actual meaning and interpretation of the perferct gas of clusters or, alternatively, of the combinatorial solution of the series expansions of the Ising model on planar lattices. We observed that the perfect gas of clusters  works for the Ising model on square and hexagonal lattices. For the triangular case,  coefficients do not have clear combinatorial interpretation. There arise several questions: \textit{(i)} What are the sufficient conditions for the graph which results in proper (i.e. without appearance of the negative probabilities) gas of clusters interpretation? \textit{(ii)} What is the meaning of those negative probabilities in the context of the lattice models? \textit{(iii)} What is the general (i.e. true for both: positive and negative cases) interpretation of the perfect gas of clusters in the context of the Ising model?

\appendix

\section{Equivalence of the free energy formulas for the triangular lattice}\label{sec:formulae}
As it was mentioned in the introductory section we have considered  free energy in the form of Eq. (127) from \cite{domb1}, but historically first was the result of Wannier \cite{wannier}. In the following section we present calculations which justify equivalence of those two formulas. Let us start with Eq. (33)\footnote{It is worth to notice that $J$ in our notation is equal to a half of Wanniers' $J$ and we write this formula in our notation.} in \cite{wannier}
\begin{align}\label{eq:wannier}
-\beta \varphi_\triangle=&\ln \left(e^{3\beta J}+e^{-\beta J}\right)+ \frac{1}{8\pi^2}\int_0^{2\pi} d\theta_1 \int_0^{2\pi}
d\theta_2 \times\\
\nonumber\times&\ln\left[1- 2\kappa+ 2\kappa\cos\theta_1+ 2\kappa\cos\theta_2+2\kappa\cos\left(\pi-\theta_1-\theta_2\right)\right],
\end{align}
where $\kappa=(e^{4\beta J}-1)/(e^{4\beta J}+1)^2$ results 
\begin{align*}
-\beta \varphi_\triangle&=\ln \left[e^{-\beta J}\left(e^{4\beta J}+1\right)\right]+ \frac{1}{8\pi^2}\int_0^{2\pi} d\theta_1 \int_0^{2\pi}
d\theta_2 \ln\bigg[\left(e^{4\beta J}+1\right)^{-2}\times\\
\times&\bigg((e^{4\beta J}+1)^2+ 2(e^{4\beta J}-1)\big(-1+ \cos\theta_1+ \cos\theta_2- \cos(\theta_1+\theta_2)\big)\bigg)\bigg]=
\end{align*}
where we use trigonometric identity $\cos(\pi-\alpha)=-\cos(\alpha)$ and factorise the terms under logarithms, which allows to simplify them as follows
\begin{align*}
&\ln \left(e^{-\beta J}\right)+\cancel{\ln\left(e^{4\beta J}+1\right)}-\cancel{\ln\left(e^{4\beta J}+1\right)}+\frac{1}{8\pi^2}\int_0^{2\pi} d\theta_1 \int_0^{2\pi}
d\theta_2\ln\bigg[e^{8\beta J}+\\&+\cancel{2e^{4\beta J}}+1-\cancel{ 2e^{4\beta J}}+2+2(e^{4\beta J}-1)\bigg(\cos\theta_1+ \cos\theta_2- \cos(\theta_1+\theta_2)\bigg)\bigg]=
\end{align*}
where we simplified the terms with the opposite signs
\begin{align*}
&\frac{1}{2}\ln \left(e^{-2\beta J}\right)+\frac{1}{8\pi^2}\int_0^{2\pi} d\theta_1 \int_0^{2\pi}
d\theta_2\times\\&\times\ln\left[e^{8\beta J}+3+2(e^{4\beta J}-1)\left(\cos\theta_1+ \cos\theta_2- \cos(\theta_1+\theta_2)\right)\right]=
\end{align*}
where we exclude the factor $1/2$ before the first logarithmic term, which allows to merge it with the integral
\begin{align*}
&\frac{1}{8\pi^2}\int_0^{2\pi} d\theta_1 \int_0^{2\pi}
d\theta_2\times\\
\times &\ln\left[4\left(\frac{e^{6\beta J}}{4}+\frac{3e^{-2\beta J}}{4}+\frac{e^{2\beta J}-e^{-2\beta J}}{2}\left(\cos\theta_1+ \cos\theta_2- \cos(\theta_1+\theta_2)\right)\right)\right]=
\end{align*}
where the terms were merged, which allows to observe that before trigonometric terms stands hyperbolic sine, which almost leads to the final formula,
\begin{align*}
&=\ln 2+\frac{1}{8\pi^2}\int_0^{2\pi} d\theta_1 \int_0^{2\pi}
d\theta_2\times\\
\times &\ln\left[\frac{e^{6\beta J}}{4}+\frac{3e^{-2\beta J}}{4}-\sinh(2\beta J)\left(-\cos\theta_1-\cos\theta_2+ \cos(\theta_1+\theta_2)\right)\right]\nonumber.
\end{align*}
Let us finally note that 
\begin{align*}
&\cosh^3(2\beta J)+\sinh^3(2\beta J)=\frac{1}{8}\left[\left(e^{2\beta J}+e^{-2\beta J}\right)^3+\left(e^{2\beta J}-e^{-2\beta J}\right)^3\right]=\\
&=\frac{1}{8}\left(e^{6\beta J}+\cancel{3e^{2\beta J}}+3e^{-2\beta J}+\cancel{e^{-6\beta J}}+e^{6\beta J}-\cancel{3e^{2\beta J}} +3e^{-2\beta J}-\cancel{e^{6\beta J}}\right)=\\
&=\frac{1}{8}\left(2e^{6\beta J}+6e^{-2\beta J} \right)=\frac{e^{6\beta J}}{4}+\frac{3e^{-2\beta J}}{4}.
\end{align*}
which leads us to the final formula for free energy
\begin{align}
-\beta \varphi_\triangle=\ln 2&+\frac{1}{8\pi^2}\int_0^{2\pi} d\theta_1 \int_0^{2\pi}
d\theta_2\ln\bigg[\cosh^3(2\beta J)+\sinh^3(2\beta J)+\nonumber\\
&-\sinh(2\beta J)\bigg(-\cos\theta_1-\cos\theta_2+ \cos(\theta_1+\theta_2)\bigg)\bigg]\label{eq:wannier2}.
\end{align}

There is one seeming difference between Eqs. (\ref{eq:phiT}) and (\ref{eq:wannier2}) - there are opposite signs before term $\cos(\theta_1)+\cos(\theta_2)$. As one can prove \cite{observation}, this do not influence the final value of the integral, because the value, for every terms in the series expansions, depends only on the following integrals
\begin{equation*}
\int_0^{2\pi}\int_0^{2\pi}\left[\pm\left(\cos(\theta_1) +\cos(\theta_2)\right)+ \cos(\theta_1+\theta_2)\right]^n d\theta_1 d\theta_2,
\end{equation*}
as one can check, this integral's values do not depend on chosen sign, which proves that Eqs. (\ref{eq:phiT}) and  (\ref{eq:wannier2}) give the same values of free energy.

\end{document}